\documentstyle[aps,multicol,eqsecnum]{revtex}
\begin{document}
\draft
\title{Phase Structures of Magnetic Impurity Models
with Two-Body Hybridization }
\author{Yue Yu and Wan-Peng Tan}
\address{Institute of Theoretical Physics, Academia Sinica, 
Beijing 100080, China} 
\maketitle
\begin{abstract}
The most general model with a magnetic impurity coupled to hybridizing and
screening channels of a conduction band is considered.
The partition function of the system is asymptotically
equivalent to that of the
multi-component kink plasma with a weak external field.
The scaling properties of the models for
finite $U$ are sketched by using the Anderson-Yuval-Hamann-Cardy
poor man's scaling theory. We point out that it is proper to include a
two-body hybridization in order to obtain correct renormalization flows.
The phase structures are studied graphically for
the general model and various reduced models. A Fermi-non-Fermi liquid
phase transition is found for all the models. We also show all possible 
phases with different finite temperature behaviors though they have the 
same Fermi liquid fixed point at low temperature. We also discuss the 
fixed point behaviors in the mixed valence state regime.

\end{abstract}

\begin{multicols}{2}

\section{ Introduction}

The normal-state properties of the Cu-O-based materials had been
understood phenomenologically on the basis of the existence of
the non-Fermi-liquid(NFL)-like groundstates \cite{And1}\cite{Var1}. 
In one dimension (1-d), it is well-known that the Luttinger liquid states 
appear universally in a wide class of interacting systems \cite{Hald1}. 
The NFL behaviors are shown
for arbitrarily weak interactions in 1-d. However, for the dimensions higher
than one, the perturbative renormalization group analysis does not
support the NFL behaviors \cite{Sh1}. This promotes the non-perturbative
approaches to be developed. In 1-d, bosonization is a powerful
non-perturbative method \cite{Matt1}. For higher dimensions, several
ways are applied. People try to generalize the 1-d
bosonization procedure to higher dimensional one. Haldane \cite{Hald2},
Houghton and Marston \cite{Hou1} provided a kind of bosonization procedures in
higher dimensions. More recently, many authors use this kind of procedure to
discuss whether the NFL behaviors appear in higher dimensions \cite{Kh1}.
The other subject that many investigations focus on is the confinement
of the Luttinger liquid behavior among the weakly interacting 1-d
Tomonaga-Luttinger chains \cite{And2}.
There still are some arguments for this subject \cite{Wen1}.
The way related to this paper is to study such models that the properties of
the models can be obtained either exactly or with the well-defined limit of
validity. For example, various single impurity models are good objects
to be studied. Research for them can be thought of as a starting
point to do the corresponding lattice models related to the
physics of the high temperature superconduction materials.

It has been known that at low
energies which are much less than the Kondo temperature,
the groundstates of simple magnetic impurity models \cite{Fri1}\cite{And3}
are Fermi liquid (FL) \cite{Wil1}\cite{Noz1}. It has recently been argued that
the generalization of those models is closely related to
the physics of the Cu-O-based materials \cite{Per1}\cite{Gia1}\cite{Zha1}
\cite{Si1}. 
It is important that a phase transition between FL and NFL
states has been found in investigations of single impurity models
because the NFL behavior can explain many anomalies of the normal
state of high temperature superconductors \cite{And1}\cite{Var1}.

The Kondo model is one of typical
impurity models. Through the bosonization
or the partition function analog, the Kondo problem can be mapped to a
one-channel resonant-level model and has been shown that the ground state
for the energy well below the Kondo temperature is of the FL behavior
\cite{Wie1}. The multi-channel
generalized model considers all of conduction
channels interact with the impurity orbital but only one channel 
hybridizes with the impurity orbital. Following the terminology used by
the authors of \cite{Per1}\cite{Gia1} we call later the 0-channel or the 
hybridizing channel and others the screening channels. 
For the finite range interactions a FL-NFL phase transition has been shown
by multiplicative renormalization group \cite{Sol1}
in \cite{Gia1} and by the poor man's scaling theory \cite{And4} in \cite{Zha1}.
Physically, as pointed out by Zhang, Yu and Su \cite{Zha1},
the phase transition is caused by a competition between two
different kinds of metallic behaviors, the NFL behavior of the multi-channel
X-ray edge problem \cite{Mah1} in the screening channels and the FL behavior
of the one-channel Kondo problem in the hybridizing channel.
A more general impurity model is the Anderson model \cite{Fri1}\cite{And3}.
Perakis {\it et al} \cite{Per1} studied the multi-channel Anderson model and
also found that there are NFL phases by using Wilson's
numerical renormalization group
method \cite{Wil1}. A similar case also occurs for a generalized Hubbard model
in infinite dimensions \cite{Si1}.

In a previous Brief Report \cite{Yu1} by one of us, co-operating with Li and
d'Amberumenil, one showed that it is necessary to introduce a hybridization of
two hybridizing electrons with opposite spin hopping onto
impurity orbital in order to keep the renormalization group's transitivity
in the multi-channel models. Here, we would like to give our full 
investigation to this issue.
We will generalize the approach used in the
particle-hole(PH) symmetric model to more general magnetic impurity
models in the spirit of the Anderson-Yuval-Hamann-Cardy (AYHC)
poor man's scaling theory \cite{And4}\cite{Car1}, which has been generalized 
to a form with symmetry breaking \cite{SK}.
We also find that in the most general model for finite $U$
a fully renormalizable model must include a hybridization of
two hybridizing electrons with opposite spin hopping onto
impurity orbital.
Because we try to describe the low energy behaviors under a low electron
density in the conduction band, we apply the bosonization procedure for
impurity models \cite{Eme1}. The Hilbert space of the system can be
divided into four sub-spaces characterized by the impurity states. The
`dipole' operators between sub-spaces are of their transitivity. This
transitivity leads to an asymptotic equivalence, via bosonization,
between the general model and a multi-component kink plasma model. This has 
been shown by Si and Kotliar in the infinite $U$ case \cite{Si1,SK}. 
On the other hand, according to the transitivity,
the `charges' of the plasma are shown to obey the
Jacobi identities, which implies that the `charges' form an adjoint
representation of some Lie algebra. The Lie algebra is shown to be
solvable. Under the PH symmetry, the dimensions of the
Lie algebra are reduced, which is closely
related to the changes of the phase structures.
The phase structures of various models can be recognized from the
corresponding renormalization group equations which are derived by
using the AYHC poor man's scaling theory 
and its symmetry breaking generalization. The phase diagrams for the models
may be drawn systematically by Cardy's graphical method \cite{Car1}. For each
model, we find that the phase in which all fugacities of the plasma
are irrelevant is the NFL phase. As
long as one of the fugacities varies from
an irrelevant to a relevant one, the FL-NFL phase transition occurs. For the 
irrelevant phase, the model at fixed point is reduced to the  X-ray 
edge problem, whose fixed point behaviors are controlled by the NFL. 
A competition between those two different type metallic states
causes the FL-NFL phase transition. Cardy's graphical 
method also shows that it is possible that there are phase
transitions at finite temperature between other phases although they have 
the same low temperature Fermi liquid fixed point. However, 
the renormalization group equations may not be valid quantitatively 
in the strongly coupling regions. We also study fixed point
behaviors in the mixed valence regime. One sees that there are four kinds
of the different fixed points. One is in the strong coupling limit 
corresponding to the FL one and another is in the weak coupling
limit which is the NFL one.  We calculate various single-particle and 
two-particle correlation functions in the weak coupling limit.
Other two are the intermediate 
coupling regimes in which there are local NFL behaviors although
the infrared orthogonality catastrophe of the X-ray edge type does not appear. 
Comparing to
the results of the infinite $U$ model \cite{SK}, we have one more 
intermediate mixed valence regime which is related to the two-body 
hybridization irrelevant.  

This paper is organized as follows. In section II, we define the
general model under the consideration and explain various reduced models.
In section III, the partition function of the theory is expressed by
the histories of
the impurity states and is mapped into that of a multi-component kink plasma.
We also briefly review the bosonization procedure for the problem.
In section IV, we study the phase structures of
the theory. In section V, the fixed point behaviors of the mixed valence states
are discussed. The last section gives our conclusion. The appendices talk about
the Lie algebra related to the `charges' of the kinks and give some comments
on the derivation of the renormalization group equations.


\section{  Single Impurity Models with Two-body Hybridization}


We begin with the derivation of the general model. Assume a single impurity
atom is located at the origin and has a localized electron orbital
$\phi_d({\bf r})$  with the local impurity level $\epsilon_{d\sigma}$.
Other electrons are in the conduction band described by the
wave function $\phi_{\bf k}({\bf r})$ and energy $\epsilon_{\bf k}$. A
general state function is the summation over all possible states,
\begin{equation}
\psi({\bf r})=\sum_{{\bf k} \sigma}\phi_{\bf k}({\bf r})X_\sigma c_{{\bf
k}\sigma}+\sum_\sigma\phi_d({\bf r}) X_{d\sigma} {d_\sigma},
\label{s11}
\end{equation}
where $X_\sigma$ and $X_{d\sigma}$ are the spin wave
functions and $\sigma$ labels spin up $\uparrow$ and down $\downarrow$.
More generally, we consider the spin wave function of impurity,
$X_{d\sigma}$ as an operator
$X_{d\sigma}=x_\sigma (1-n_d)+{\tilde x}_\sigma
n_d$, which leads to the different wave functions for the empty,
single occupation and double occupation of the impurity orbital. Under this
consideration, the Fano-Anderson model is generalized as
\begin{eqnarray}
\displaystyle
H_{GFA}&=&\sum_{{\bf k},\sigma}\epsilon_{\bf k}c^\dagger_{{\bf k}\sigma}
c_{{\bf k}\sigma}
+\sum_\sigma \epsilon_{d\sigma}{d^\dagger_\sigma}
{d_\sigma} \label{s12}\\ \nonumber 
&+&\frac{1}{\sqrt{L}}\sum_{{\bf
k},\sigma} [(t_{{\bf k}\sigma}(1-n_d)+\tilde t_{{\bf k}
\sigma}n_d)c^\dagger_{{\bf k}\sigma}{d_\sigma} +h.c.],
\end{eqnarray}
where $\epsilon_{d\sigma}$ is the impurity level which is taken as
the spin independence, $\epsilon_d$,
without losing the generality. Assume only the conduction electrons 
in the 0-channel hybridize with the
impurity orbital and the hybridizations are $k$-independent in the model.
The chemical potential is set to zero but the
Hamiltonian (\ref{s12}) is still particle-hole-asymmetry.

To discuss the interesting physical phenomenon, the electron-electron
interaction must be introduced. We consider only the finite range
interaction. Besides the Hubbard-type interaction in the Anderson model
we also keep consideration to other two types of interactions,
the spin-exchange interaction and a two-body hybridization
which describes the two hybridizing electrons with opposite spins
hopping onto the impurity orbital simultaneously.
By using the wave function (\ref{s11}),  the model Hamiltonian reads
\begin{eqnarray}\displaystyle
H&=&\displaystyle\sum_{k>0;\sigma,l}\epsilon_k c^\dagger_{k\sigma l}c_{k\sigma 
l}+
\epsilon_dn_d+Un_{d\uparrow}n_{d\downarrow} \label{s13} \\ \nonumber
&+&\displaystyle\frac{1}{\sqrt{L}}\sum_\sigma[(t_{1\sigma}(1-n_d)+\tilde
t_{1\sigma}n_d)c^\dagger
_{\sigma 0}d_\sigma+h.c.]\\ \nonumber
&+&\frac{t_2}{L}\sum_\sigma(c^\dagger_{\sigma
0}c^\dagger_{-\sigma 0}d_{\sigma}d_{-\sigma}
+h.c.)\\ \nonumber
&+&\displaystyle\frac{1}{L}\sum_{\sigma l}[V_{x\sigma l}(2-n_d)+\tilde
V_{x\sigma l}(n_d-1)]c^\dagger_{\sigma l}
c_{\sigma l}d^\dagger_\sigma d_\sigma\\ \nonumber
&+&\frac{1}{L}\sum_{\sigma
l}V_{yl}c^\dagger_{\sigma
l}c_{-\sigma l}d^\dagger_{-\sigma}d_\sigma\\ \nonumber
&+&\displaystyle\frac{1}{L}\sum_{\sigma\sigma'
l}[V_{l\sigma}(2-n_d)+\tilde V_{l\sigma}(n_d-1)]c^\dagger_{\sigma l}
c_{\sigma l}d^\dagger_{\sigma'}d_{\sigma'},
\end{eqnarray}
The first four terms in (\ref{s13}) are the generalized Fano-Anderson model 
(\ref{s12})
added by the Hubbard interaction, which is back to the Anderson model
if $t_{1\sigma}=\tilde t_{1\sigma}\equiv t_1$. In the meanwhile,
$V_{l\sigma }$
term takes into account the finite range interactions between the
impurity orbital and conduction electrons ($l=0,...,N_f$). Without losing
the generality, we consider the screening channels as the spinless fermions
and take
$V_{l\sigma}=\tilde
V_{l\sigma }=V_l$ for $l=1,..,N_f$ for convenience.
The $V_{x\sigma l}$ and $V_{yl}$ terms describe the spin exchange interactions.
We take  those with $l=1,...,N_f$ vanishing for convenience.
The $t_2$ term which is newly introduced in the Hamiltonian
represents that two hybridizing channel electrons with opposite 
spins hop onto the impurity orbital simultaneously.
To renormalize the model it is proper that this term
as well as spin exchange terms are included in the Hamiltonian.

In (\ref{s13}), if the coupling constants with the `tilde' are equal to those
without the `tilde', the Hamiltonian is reduced to the PH symmetric one
under $\epsilon_d=-U/2$. As we have shown that for this reduced model
the consistency of the renormalization group equations
forces the PH symmetry have to hold
\cite{Yu1}. This Hamiltonian  also includes the
Kondo model as a special case \cite{Yu1}.
Spinless models can also be reached by  a proper choice of the coupling
constants.

We shall show there is a FL-NFL phase transition for the system.
To understand the physical mechanics
of the appearance of the phase transition, we
separate the Hamiltonian (\ref{s13}) into two parts, as done by Zhang, 
Yu and Su \cite{Zha1} for the spinless impurity model, $H=H_h+H_s$, where
\begin{eqnarray}\displaystyle
H_h&=&\displaystyle\sum_{k>0,\sigma}\epsilon_kc^\dagger_{k\sigma } c_{k\sigma }
+\epsilon_d n_d+Un_{d\uparrow}n_{d\downarrow} \label{s14} \\ \nonumber 
&+&\displaystyle\frac{1}{\sqrt{L}}\sum_\sigma[(t_{1\sigma}(1-n_d)+\tilde
t_{1\sigma}
n_d)c^\dagger_{\sigma }d_\sigma+h.c.]\\ \nonumber
&+&\frac{t_2}{L}\sum_\sigma (c^\dagger_{\sigma }c^\dagger_{-\sigma }d_\sigma
d_{-\sigma}+h.c.)
\\ \nonumber
&+&\displaystyle\frac{1}{L}
\sum_{\sigma}[V_{x\sigma }(2-n_d)+\tilde V_{x\sigma}(n_d-1)]c^\dagger
_{\sigma }c_{\sigma }d^\dagger_\sigma d_\sigma \\  \nonumber &+&
\frac{1}{L}\sum_{\sigma}
V_{y}c^\dagger_{\sigma }c_{-\sigma}d^\dagger_{-\sigma}d_\sigma
\\ \nonumber &+& \displaystyle
\frac{1}{L}\sum_{\sigma,\sigma'}[V_{0\sigma}(2-n_d)+\tilde V_{0\sigma}(n_d-1)]
c^\dagger_{\sigma }c_{\sigma } d^\dagger_{\sigma'}d_{\sigma'},\\
\displaystyle
H_s&=&\displaystyle\sum_{k,l>0}\epsilon_k c^\dagger_{k l}c_{k l}
+\frac{1}{L}\sum_{l>0}V_lc^\dagger_{ l}c_{ l}n_d.
\label{s15}
\end{eqnarray}
Here we have suppressed most of the suffices `0'. It is easy to see that the
screening part, $H_s$, of $H$ is the multi-channel X-ray edge model,
whose properties are controlled by a NFL fixed point; while the hybridizing
part, $H_h$, of $H$ is a one-channel impurity model, whose properties are
given by some FL fixed points with different symmetries.
Since these two parts are coupled by the interactions between
the conduction band and the impurity orbital, there appears
an interesting competition phenomenon between FL and NFL metallic
behaviors, which causes the phase transition.


\section{ Partition Function}


In this section we show that the partition function of the model at low 
energies may be equivalent to that of the multi-component kink plasma model 
with a weak external field.

\subsection{ Bosonization and Sum over Impurity Histories}

Since we deal with the low density system,
the spectrum of the conduction band can be taken as the linear form
$\epsilon_k=(k-k_f)/\rho$ with $\rho=(hv_F)^{-1}$ being the density
of state at the Fermi surface. For the linear spectrum the conduction
band can be treated by the bosonization technique \cite{Matt1}\cite{Eme1}. 
We first bosonize the screening part of the model, $H_s$. We introduce the
density operators:
$$              \displaystyle
\rho_{l}(k)=\frac{1}{\sqrt L}\sum_{q=0}^{W-k}c^\dagger_{q
l} c_{q+k ,l} ~~~,
~
\rho_{l}(-k)=\frac{1}{\sqrt L}\sum_{q=k}^{W}c^\dagger_{q
l}c_{q-k ,l},
$$
where $W=1/\tau_0$ is the conduction band width which is much greater than
$U$ and $\epsilon_d$. The operators obey the Bose commutation relations
$$
[\rho_{l}(k),\rho_{ l'}(-k')]=k\delta_{k,k'} \delta_{l,l'},
$$
 accordance with which,
we can define standard bosonic creation and annihilation
operators
$$
b_{k l}=-i\sqrt{\frac{1}{k}}\rho_{ l}(k)~~~,~~~
b^\dagger_{kl}=i\sqrt{\frac{1}{k}}\rho_{l}(-k).
$$
The screening electron operator is  represented, via the bosonic operator,
as
\begin{eqnarray}
&&c^\dagger_{l}(x)=\frac{1}{\sqrt{\tau_0}}\exp[-i\phi_l(x)],
\\ \nonumber \displaystyle
&&\phi_l(x)=\sum\sqrt{\frac{1}{kL}} (b^\dagger_{kl}e^{-ikx}+b_{kl}e^{ikx}),
\label{5}
\end{eqnarray}
and the screening Hamiltonian, $H_s$, is transformed to
\begin{equation}
H_s=\sum_{k,l>0}\frac{k}{\rho}b^\dagger_{kl}b_{kl}
-\sum_{k,l>0}\frac{i V_l}{\sqrt{L}}\sqrt{k}(b^\dagger_{k l}-b_{k l})n_d,
\label{6}
\end{equation}
which can be diagonalized by the following canonical transformation
$$
T=\exp\{i\sum_{k,l>0}\frac{\rho V_l}{\sqrt{k L}}(b^\dagger_{k l}
+b_{k l})n_d\}.
$$
Since this canonical transformation does not commute with $d_\sigma$, the
hybridization terms are transformed to the coupling forms among the impurity,
hybridizing and screening channels while other terms do not change in $H_h$.
We have the effective Hamiltonian, $H=H_0+H_I$, with
\begin{eqnarray}\displaystyle
H_0&=&\displaystyle\sum_{k>0,\sigma}\epsilon_kc^\dagger_{k\sigma} c_{k\sigma}
+\epsilon_dn_d+U~n_{d\uparrow}n_{d\downarrow} \nonumber \\ \nonumber
&+&\displaystyle\frac{1}{L}\sum_{\sigma,\sigma'}[V_{0\sigma}(2-n_d)+\tilde
V_{0\sigma}(n_d-1)]c^\dagger_\sigma
c_\sigma d^\dagger_{\sigma'}d_{\sigma'}\\  \nonumber
&+&\displaystyle\frac{1}{L}\sum_\sigma[V_{x\sigma}(2-n_d)
+\tilde V_{x \sigma}(n_d-1)] c^\dagger_\sigma
c_\sigma d^\dagger_\sigma d_\sigma \\ \label{s27}
&+&\sum_{k,l>0}\frac{k}{\rho}
b^\dagger_{kl}b_{kl},\\ \nonumber
H_I&=&\displaystyle
\frac{1}{\sqrt{L}}\sum_\sigma[(t_{1\sigma}(1-n_d)+\tilde
t_{1\sigma}n_d)\Delta^\dagger
c^\dagger_\sigma d_\sigma
+h.c.]\\ \nonumber
&+&\frac{t_2}{L}\sum_{\sigma}(\Delta^{\dagger 2}c^\dagger_\sigma
c^\dagger_{-\sigma}d_\sigma d_{-\sigma}+h.c.) \\ \nonumber
&+&\displaystyle\frac{V_y}{L}\sum_\sigma
c^\dagger_\sigma c_{-\sigma}d^\dagger_{-\sigma}d_\sigma.
\end{eqnarray}
The operator
$\Delta^\dagger$ is given by
$\Delta^\dagger=\exp\{\sum_{k,l>0}\displaystyle\frac{i\rho V_l}{\sqrt{k L}}(
b^\dagger_{k l}+b_{k l})\}$  or in the coordinate representation
$\Delta^\dagger(x)=\exp\{i\sum\displaystyle\frac{\delta_l}{\pi}\phi_l(x)\}$ 
where $\delta_l=\pi\rho V_l$.

Now, we turn to derive the partition function of the Hamiltonian (\ref{s27}), 
where $H_0$ and $H_I$ are
considered as the unperturbed Hamiltonian  and the
interaction Hamiltonian, respectively. Hence, we have a partition function
\begin{equation}
Z=Tr(e^{-\beta H})=Tr(e^{-\beta H_0}T_\tau \exp\{-\int_0^\beta H_I(\tau)d\tau
\}),
\label{8}
\end{equation}
where $\beta$ is the inverse of the temperature, $T_\tau$ is the (imaginary)
time ordering symbol and $H_I(\tau)=e^{H_0\tau}H_Ie^{-H_0\tau}$ is the
interaction-picture operator of $H_I$. Setting a sequence of times $0<\tau_1
<...<\tau_n<\beta$, the partition function is expanded as $Z=\displaystyle\sum
_{n=o}^\infty Z_n$, where
\begin{equation}
Z_n=(-1)^n\int_0^\beta d\tau_n...\int_0^{\tau_2} d\tau_1 Tr
(e^{-\beta H_0}H_I(\tau_n)...H_I(\tau_1)).
\label{9}
\end{equation}

The Hilbert space of $H_0$  can be projected to four sub-spaces
characterized by four impurity electron states, $|\alpha>=|0>, |\sigma>$ and
$|3>\equiv|\uparrow\downarrow>$. Each term in the interaction $H_I$ plays 
a role of the `dipole' operator causing a transition between
those sub-spaces. Such a kind
of problems can be treated in Haldane's familiar way \cite{Hald3}:
Write the partition function in terms of a sum
over histories of the impurity. We will follow the formalism
given by Si and Kotliar \cite{Si1,SK}. Setting $P_\alpha=|\alpha><\alpha|$
is a projection
operator to $\alpha$-state subspace, $H_0$, then, is projected as
\begin{equation}
H_0=\sum_\alpha H_\alpha P_\alpha ,
\label{10}
\end{equation}
where
\begin{eqnarray}
H_\alpha&=&\epsilon_{d,\alpha}+U_\alpha+\sum_\sigma V^\sigma_\alpha
c^\dagger_\sigma c_\sigma \label{s211}\\ \nonumber
&+&\sum_{k>0;\sigma}\epsilon_k c^\dagger_{k\sigma}
c_{k\sigma}+\sum_{k,l>0}\frac{k}{\rho}b^\dagger_{k l}b_{k l}.
\end{eqnarray}
The impurity levels, here, are
projected as $\epsilon_{d,0}=0$, $\epsilon_{d,\sigma}=
\epsilon_d$ and $\epsilon_{d,3}=2\epsilon_d$; the Hubbard interactions are
$U_0=U_\sigma=0$ and $U_3=U$. Making a shift of the ground state
energy, we can redefine $\epsilon_{d,\alpha}+U_\alpha$ as 
$\varepsilon_\alpha$
with $\varepsilon_0=-\epsilon_d-U/4$, $\varepsilon_\sigma=-U/4$ and
$\varepsilon_3=\epsilon_d+3U/4 $, which obey $\sum \varepsilon_\alpha=0$.
The potentials are given by
$$
V^\sigma_0=0,~~V^\sigma_{\sigma'}=\frac{V_{0\sigma}}{L}+\frac{V_{x\sigma}}{L}
\delta_{\sigma\sigma'},~~V^\sigma_3=\frac{2\tilde V_{0\sigma}}{L}+\frac{\tilde V
_{x\sigma}}{L}.
$$

Now, we sum over the impurity states by inserting a complete set
of the impurity states at every discrete time. Thus, (\ref{9}) is rewritten as
\begin{eqnarray}
Z_n&=&\displaystyle(-1)^n\sum_{\alpha_1, \alpha_2,...,\alpha_n}
\int_0^\beta d\tau_n...\int_o^{\tau_2}d\tau_1\\ \nonumber
&\times&
g'_{\alpha_{n+1}\alpha_n}...g'_{\alpha_2\alpha_1} \\ \nonumber
&\times&
tr_c\{e^{-H_{\alpha_{n+1}}(\beta-\tau_n)}O'(\alpha_{n+1},\alpha_n)e^{-H_{
\alpha_n}(\tau_n-\tau_{n-1})}\\ \nonumber
&&...e^{-H_{\alpha_2}(\tau_2-\tau_1)}O'(\alpha_2,
\alpha_1)e^{-H_{\alpha_1}}\},
\end{eqnarray}
where $tr_c$ indicates the trace carried out for the conduction band.
$\alpha_1,...,\alpha_n$ and $\tau_1,...,\tau_n$
label a Feynman trajectory with $\alpha_{n+1}=\alpha_1$.
$g'_{\alpha\beta}$ are coupling constants,
$g'_{0\sigma}(g'_{-\sigma 3})\equiv g'_{1\sigma}
(\tilde g'_{1\sigma})=t_{1\sigma}/\sqrt{L}(\tilde
t_{1\sigma}/\sqrt{L})$,
$g'_{\sigma,-\sigma}
\equiv g'_y=V_y/L$ and $g'_{03}\equiv g'_{2}=t_2/L$. The `dipole' operators
$O'(\alpha,\beta)$ are given by
\begin{eqnarray} \displaystyle
&&O'(0,\sigma)={O'}^\dagger(\sigma,0)\\ \nonumber
&&=O'(-\sigma,3)=
{O'}^\dagger(3,-\sigma)=\Delta^\dagger c^\dagger_\sigma,\\ 
\nonumber \displaystyle
&&O'(0,3)={O'}^\dagger(3,0)=\Delta^{\dagger 2}c^\dagger_\uparrow
c^\dagger_\downarrow ,\\ \nonumber
&&O'(\sigma,-\sigma)= c^\dagger_{-\sigma}c_\sigma.
\end{eqnarray}

A similar bosonization procedure to the screening channels can be applied
to the impurity channel. As a result, the projected Hamiltonian transforms
into
\begin{eqnarray}
H_\alpha=\displaystyle
H_c+\varepsilon_\alpha-\sum_{k>0;\sigma}\frac{i\delta_\alpha^\sigma}
{\pi \rho}\sqrt{\frac{k}{L}}(b^\dagger_{k\sigma}-b_{k\sigma}), \label{s214}\\ 
\nonumber
\displaystyle H_c=\sum_{k>0;\rho}\frac{k}{\rho}b^\dagger_{k\sigma}b_{k\sigma}
+\sum_{k,l>0}\frac{k}{\rho}b^\dagger_{kl}b_{kl},
\end{eqnarray}
where $b_{k\sigma}$, the bosonic operator corresponding to $c_{k\sigma}$,
is given through
\begin{eqnarray}
&&c^\dagger_\sigma(x)=\exp[-i\phi_\sigma(x)],\\ \nonumber
&&\displaystyle\phi_\sigma(x)=\sum_k\sqrt{\frac{1}{kL}}(b^\dagger_{k\sigma}e^{-i
kx}+
b_{k\sigma}e^{ikx}).
\end{eqnarray}
The phase shifts $\delta^\sigma_\alpha=\pi\rho V^\sigma_\alpha L$ are defined by
\begin{equation}
\delta_0^\sigma=0,~~~\delta_{\sigma'}^\sigma=\delta_0+\delta_{x\sigma}
\delta_{\sigma'\sigma},~~~\delta_3^\sigma
=2\tilde\delta_0+\tilde\delta_{x\sigma},
\end{equation}
where $\delta_0=\displaystyle\pi\rho V_{0\sigma}$, $\tilde\delta_0=
\displaystyle\pi\rho \tilde V_{0\sigma}$, $\delta_{x\sigma}=\displaystyle
\pi\rho V_{x\sigma}$ and $\tilde\delta_{x\sigma}=\displaystyle
\pi\rho \tilde V_{x\sigma}$ are the Born approximation values
of the general phase shifts $2\tan^{-1}\displaystyle(\pi\rho V_{0\sigma}/2)$ , 
and so 
on.
The Hamiltonian (\ref{s214}) can also be diagonalized to
$H_c+\varepsilon_\alpha$ by the following canonical transformations
\begin{equation}
U_\alpha=\Pi_\sigma U_{\delta_\alpha^\sigma}=\Pi_\sigma
\exp\{i\delta_\alpha^\sigma \phi_\sigma\}.
\end{equation}
Meanwhile, the bosonized `dipole' operators can be transformed to
\begin{equation}
O(\alpha,\beta)=exp\{iq^a_{\alpha\beta}\phi_a\},
\label{s218}
\end{equation}
where $\phi_a=(\phi_\sigma,\phi_l)$. The quantities $q^a_{\alpha\beta}$, which
will be called `charges', are
\begin{eqnarray}\displaystyle
q^a_{0\sigma}&=&\biggl((\frac{\delta_{x\sigma'}}{\pi}-1)
\delta_{\sigma\sigma'}+\frac{\delta_0}{\pi}, \frac{\delta_l}{\pi}\biggr),
\\ \nonumber \displaystyle
q^a_{\sigma-\sigma}&=&\biggl((\frac{\delta_{x\sigma'}}{\pi}-1)
(\delta_{\sigma'-\sigma}-\delta_{\sigma'\sigma}),0\biggr),\\ \nonumber
 \displaystyle
q^a_{-\sigma 3}&=&\biggl((\frac{\delta_{x\sigma'}}{\pi}-1)
\delta_{\sigma',\sigma}
+\frac{\delta_0}{\pi}\\ \nonumber
&+&\frac{1}{\pi}(\tilde\delta_{x\sigma'}-\delta_{x\sigma'}
+2\tilde\delta_0-2\delta_0),\frac{\delta_l}{\pi}\biggr),\\ \nonumber
\displaystyle
q^a_{03}&=&\biggl(\frac{\tilde\delta_{x\sigma'}+2\tilde\delta_0}{\pi}-1,
\frac{2\delta_l}{\pi}\biggr),
\end{eqnarray}
where $a=(\sigma',l)$
and the `charges' are anti-symmetric: $q^a_{\alpha\beta}=-q^a_{\beta\alpha}$. 
Using
the bosonic `dipole' operators (\ref{s218}) and defining the 
interaction-picture 
operators
$O(\alpha,\beta)(\tau)\equiv e^{H_c\tau}O(\alpha,\beta)e^{-H_c\tau}$, the
partition function reads
\begin{eqnarray} \displaystyle
Z_n&=&\displaystyle(-1)^n\sum_{\alpha_1,...,\alpha_n}\int_0^{\beta
-\tau_0}\frac{d\tau_n}{\tau_0}...\int_0^{\tau_2-\tau_0}\frac{d\tau_1}{\tau_0}
\label{s220}\\ \nonumber
&\times&\displaystyle g_{\alpha_{n+1}\alpha_n}...g_{\alpha_2\alpha_1}
\exp\{-\sum_i E_{\alpha_i+1}\frac{\tau_{i+1}-\tau_i}{\tau_0}\}\\ \nonumber
&\times&tr_c(e^{
-H_c\beta}O(\alpha_{n+1},\alpha_n)(\tau_n)...
O(\alpha_2,\alpha_1)(\tau_1)),
\end{eqnarray}
where $g_{\alpha\beta}=g'_{\alpha\beta}\tau_0$ are dimensionless coupling
constants and $E_\alpha=\varepsilon_\alpha\tau_0$ are regarded as
dimensionless external fields. The trace part of (\ref{s220}) is just the 
correlation
function of $n$-vertex operators. In the zero temperature limit, it can be
calculated by using the free boson field theory. The result is given rise to:
\begin{eqnarray}
Z&=&\displaystyle\sum_{n}\sum_{\alpha_1,...,\alpha_n}\int_0^{\beta-\tau_0}
\frac{d\tau_n}{\tau_0}...\int_0^{\tau_2-\tau_0}\frac{d\tau_1}{\tau_0} 
\label{s221}\\ 
\nonumber
&\times&\exp(-S[\tau_1,
...,\tau_n;\alpha_1,...,\alpha_n]),\\ 
S[\tau, 
\alpha]&=&\displaystyle\sum_{i<j}\sum_{a=\sigma,l}q^a_{\alpha_i\alpha_{i+1}}
q^a_{\alpha_j\alpha_{j+1}}\ln\frac{\tau_j-\tau_i}{\tau_0}\\ \nonumber
&-&\sum_i\ln(g_{\alpha_{i
+1}
\alpha_i})+\sum_iE_{\alpha_{i+1}}\frac{\tau_{i+1}-\tau_i}{\tau_0},
\end{eqnarray}
which may be regarded to describe
a one dimensional multi-component plasma of kinks with `charges' $q^a_{\alpha
\beta}$ and `fugacities' $g_{\alpha\beta}$. $E_\alpha$ is an `external field'.

\subsection{ Related to Charge-Spin
Chain and Transitivity of `Dipole' operators}

The partition function (\ref{s221}) describes a multi-component kink plasma 
with logarithmic interactions. There are six types of kinks which
correspond to the different pairs of the impurity states, the charge pairs
$(0\sigma), (\sigma 3)$ and $(03)$ and the spin pair $(\uparrow\downarrow)$.
This leads to the logarithmic interactions between the kink-pairs can be
transformed to the interactions among the charge- and spin-pairs. To reach
this point we note that the `charges' of kinks are of their transitivity:
\begin{equation}
q^a_{\alpha\beta}+q^a_{\beta\gamma}=q^a_{\alpha\gamma}.
\label{s222}
\end{equation}
This leads to the rewriting of the action $S[\tau,\alpha]$, i. e.
\begin{eqnarray}
S[\tau, \alpha]&=&\displaystyle\sum_{i<j}\tilde K(\alpha_i,\alpha_{i+1};
\alpha_j
,\alpha_{j+1})\ln\frac{\tau_j-\tau_i}{\tau_0}
\label{s223}\\ \nonumber
&-&\sum_i\ln(g_{\alpha_{i+1}\alpha
_i})+\sum_iE_{\alpha_{i+1}}\frac{\tau_{i+1}-\tau_i}{\tau_0},
\end{eqnarray}
where
\begin{eqnarray}
&&\tilde K(\alpha_i, \alpha_{i+1};\alpha_j,\alpha_{j+1})=K(\alpha_i,\alpha_j)
\\ \nonumber&&+
K(\alpha_{i+1},\alpha_{j+1})-K(\alpha_i,\alpha_{j+1})-K(\alpha_{i+1},\alpha_j)
,\end{eqnarray}
where $K(\alpha,\beta)$, called charge- and spin- stiffness constants, are
defined by \cite{29}
\begin{equation}
K(\alpha,\beta)=-\frac{1}{2}\sum_a(q^a_{\alpha\beta})^2.
\label{s225}
\end{equation}
The model described by (\ref{s223}) is
equivalent to the one-dimensional charge-spin chain model with
$1/\tau^2$ interaction \cite{Car1} and incorporating the effective fields
\cite{SK}:
\begin{eqnarray}
S[\alpha,\tau]&=&\sum_{i<j}K(\alpha_i,\alpha_j)\frac{\tau_0^2}{(\tau_j-\tau_i)
^2}-\sum_i \ln(g_{\alpha_{i+1}\alpha_i})\\\nonumber
&+&\sum_i E_{\alpha_{i+1}}\frac
{\tau_{i+1}-\tau_i}{\tau_0}
\end{eqnarray}
The model is just a special case of the general one-dimensional model
with $1/r^2$ interaction considered by Cardy \cite{Car1}. It is worth to
point out that the transitivity (\ref{s222}) plays an important role
to reach  Cardy's model. In fact, the transitivity of the `charges'
stems from the transitivity of the `dipole' operators:
\begin{equation}
O(\alpha,\beta)O(\beta,\gamma)=O(\alpha,\gamma).
\label{s227}
\end{equation}
This transitivity implies that all `dipole' operators together with the
vanishing operator $\phi$ and the identity operator $I$
form a semigroup under the operation:
Either $O(\alpha,\beta)O(\gamma,\delta)=\phi$
for $\beta\not=\gamma$ or the transitivity (\ref{s227}). Physically,
the semigroup relates to the transitivity of the charge-spin-flips:
{\it e.g.}, the pair $(03)$ can be separated as two pairs $(0\sigma)$ and
$(\sigma 3)$, {\it etc}. If the transitivity of the charge-spin-flips is
broken, the charge-spin chain is interrupted. Therefore, the model is not
self-consistent. Originally, the transitivity comes from the two-body
hybridization and the spin-exchange interaction in the Hamiltonian (\ref{s13}).
Therefore, the inclusion of them in (\ref{s13}) is proper to the 
self-consistency
of the model. In the next section, we shall see that the correct
renormalization flows mustn't exclude those of
the fugacities $g_{\uparrow\downarrow}$ and $g_{03}$.

To conclude this section, we would like to mention that the
transitivity of the `charges' and their anti-symmetry leads to
Jacobi identities, which implies the existence of a Lie algebra.
We arrange an appendix ( appendix A)
to  discuss  the properties of the Lie algebra.

\section{ Phase Structures}

In this section, we study the phase structures of the models
by using the AYHC poor man's scaling theory. The details of
the derivation of the renormalization group equations have been given in
Appendix B. Using those equations, we analyze the phase structures of
the general model and various reduced models.

\subsection{ General Model}

In the most general case, six non-zero fugacities $g_{\alpha\beta}$ are
independent of each other. Furthermore, six stiffness constants
are the functions of six phase shifts $\delta_0$,
$\tilde\delta_0$, $\delta_{x\sigma}$ and $\tilde\delta_{x\sigma}$ in the 
quadrate.
The six stiffness constants are also independent of each other.
To see this point, we consider the small phase shifts. In this
case (\ref{s225}) expresses a non-singular linear transformation if the 
quadrate
are neglected. Thus, the renormalization group equations are exactly
the 4-state example of the Cardy's general results \cite{Car1} and their
symmetry breaking generalization \cite{SK} (also see
Appendix B):
\begin{eqnarray}
\displaystyle
\frac{dg_{1\sigma}}{dln\tau_0}&=&(1-K_{0\sigma})g_{1\sigma}+g_{1,-\sigma}g_y
e^{-E_{-\sigma}+\frac{1}{2}(E_0+E_\sigma)}  \label{s31}\\ \nonumber
&+&g_2\tilde g_{1-\sigma}
e^{-E_3+\frac{1}{2}(E_0+E_\sigma)},\\ \nonumber
\displaystyle
\frac{dg_y}{dln\tau_0}&=&(1-K_{\sigma,-\sigma})g_y+g _{1\sigma}g_{1,-\sigma}
e^{-E_0+\frac{1}{2}(E_0+E_{-\sigma})}\\ \nonumber
&+&
\tilde g_{1\sigma}\tilde g_{1,-\sigma}
e^{-E_3+\frac{1}{2}(E_0+E_{-\sigma})},\\ \nonumber
\displaystyle\frac{d\tilde g_{1\sigma}}{dln\tau_0}&=&
(1-K_{-\sigma 3})\tilde g_{1\sigma}
+g_{1-\sigma}g_2e^{-E_0+\frac{1}{2}(E_3+E_{-\sigma})}\\ \nonumber 
&+&g_y\tilde 
g_{1,-\sigma}e^{-E_{\sigma}+\frac{1}{2}(E_3+E_{-\sigma})}, \\ \nonumber
\displaystyle\frac{dg_2}{dln\tau_0}&=&
(1-K_{03})g_2+\sum_\sigma\displaystyle g_{1\sigma}\tilde 
g_{1-\sigma}
e^{-E_\sigma+\frac{1}{2}(E_0+E_{3})},
\end{eqnarray}
for the fugacities. Here $K_{\alpha\beta}=-K(\alpha,\beta)$. For the
external fields, i. e. $U$ and $\epsilon_d$, one has
\begin{eqnarray}
\displaystyle\tau_0\frac{d\epsilon_d}{dln\tau_0}
&=&\displaystyle \tau_0\epsilon_d+(g^2_{1\uparrow}+g^2_{1\downarrow})
e^{E_0-E_\uparrow} \label{s32}\\ \nonumber
&-&\frac{1}{2}(g^2_{1\uparrow}+g^2_{1\downarrow})e^{E_\uparrow-E_0}\\
\nonumber
&-&\displaystyle \frac{1}{2}(
\tilde g^2_{1\uparrow}+\tilde g^2_{1\downarrow})e^{E_\uparrow-E_3}
+g^2_2 e^{E_0-E_3}-g^2_y,\\ \nonumber
\displaystyle\tau_0\frac{dU}{dln\tau_0}
&=&\displaystyle \tau_0 U+(g^2_{1\uparrow}+g^2_{1\downarrow})
(e^{E_\uparrow-E_0}-e^{E_0-E_\uparrow})+2g^2_y \\ \nonumber
&+&\displaystyle (\tilde g^2_{1\uparrow}+\tilde g^2_{1\downarrow})
(e^{E_\uparrow-E_3}-e^{E_3-E_\uparrow})\\ \nonumber
&-&g_2^2(e^{E_0-E_3}+e^{E_3-E_0}).
\end{eqnarray}
For the stiffness constants, one has
\begin{eqnarray}
\frac{dK_{\alpha\beta}}{dln\tau_0}&=&-\sum_\gamma g^2_{\alpha\gamma}
e^{E_\alpha-E_\gamma}(K_{\alpha
\beta}+K_{\alpha\gamma}-K_{\beta\gamma})
\label{s33}\\ \nonumber &-&\sum_\gamma g^2_{\beta\gamma}
e^{E_\beta-E_\gamma}(K_{\alpha\beta}+K_{\beta\gamma}-K_{\alpha\gamma}).
\end{eqnarray}
We make a few comments on the renormalization group equations (\ref{s31}), 
(\ref{s32}) and (\ref{s33}).

(i) As we have seen, if the model can be mapped to the special model
of Cardy's, $V_y$ and $t_2$ can not vanish. Furthermore, from (\ref{s31}), 
we see that even
we start from the vanishing initial values of $g_2$ and $g_y$, they will
increase as $g_{1\sigma}$ and $\tilde g_{1\sigma}$. This means that the flows
of the vanishing values of them are not correct renormalization flows.

(ii) Since our renormalization group is perturbative in its treatment of
$g_{\alpha\beta}$, the phase structure of the system can be described by
$g_{\alpha\beta}$.
The strong-coupling fixed points in the quantum impurity problem
are outside the range of validity of the AYHC poor man's scaling theory,
but it can give the correct flow directions. In the strongly coupling
limit, the fixed point behaviors need to be more carefully discussed.

(iii) Near the fixed point, $\epsilon_d, U \rightarrow 0$ in the 
order of $g^2_{\alpha\beta}$. Therefore, we can regard all $e^{E_\alpha}=1$
when we discuss the behaviors near the fixed point. Then, we can rewrite
the renormalization group equations in the form of perturbative 
$g_{\alpha\beta}$ as follows:
\begin{eqnarray}
\displaystyle
\frac{dg_{1\sigma}}{dln\tau_0}&=&(1-K_{0\sigma})g_{1\sigma}+g_{1,-\sigma}g_y
+g_2\tilde g_{1-\sigma}, 
\label{s331}\\ \nonumber
\displaystyle
\frac{dg_y}{dln\tau_0}&=&(1-K_{\sigma,-\sigma})g_y+g _{1\sigma}g_{1,-\sigma}+
\tilde g_{1\sigma}\tilde g_{1,-\sigma},\\ \nonumber
\displaystyle\frac{d\tilde g_{1\sigma}}{dln\tau_0}&=&(1-K_{-\sigma 3})\tilde 
g_{1\sigma}
+g_{1-\sigma}g_2+g_y\tilde g_{1,-\sigma}, \\ \nonumber
\displaystyle\frac{dg_2}{dln\tau_0}&=&(1-K_{03})g_2+\sum_\sigma\displaystyle 
g_{1\sigma}\tilde 
g_{1-\sigma},
\end{eqnarray}

\begin{eqnarray}
\displaystyle\tau_0\frac{d\epsilon_d}{dln\tau_0}
&=&\displaystyle \tau_0\epsilon_d+\frac{1}{2}(g^2_{1\uparrow}+g^2_{1\downarrow}
-\tilde g^2_{1\uparrow} \label{s332}\\ \nonumber
&-&\tilde g^2_{1\downarrow})+g^2_2-g^2_y,\\ \nonumber 
\displaystyle\tau_0\frac{dU}{dln\tau_0}&=&\displaystyle \tau_0 U+2(g^2_y-g_2^2).
\end{eqnarray} 

\begin{eqnarray}
\frac{dK_{\alpha\beta}}{dln\tau_0}&=&-\sum_\gamma g^2_{\alpha\gamma}(K_{\alpha
\beta}+K_{\alpha\gamma}\\ \nonumber
&-&K_{\beta\gamma})-\sum_\gamma g^2_{\beta\gamma}
(K_{\alpha\beta}+K_{\beta\gamma}-K_{\alpha\gamma}).
\end{eqnarray}

Under those considerations and according to (\ref{s331}), we can describe the 
phase structure of the system.

(i) For all $K_{\alpha\beta}>1$, all fugacities are irrelevant and renormalized
to zero. There exists a weak coupling fixed point $g^*_{\alpha\beta}=0$.
The fixed point Hamiltonian is similar to that of the multi-channel X-ray edge
problem \cite{Mah1}. The system exhibits a power law decay of the correlation
function with a non-universal exponent. This is a NFL phase.
By the definitions of $K_{0\sigma}$, the existence of the screening channels
ensures the possibility of $K_{0\sigma}>1$ and then reaches the NFL phase.
The one-channel model in a small exchange interaction can not
have this phase \cite{Zha1}. 

(ii) Once one of the coupling constant becomes relevant, the low temperature
fixed point of the system, in general, is Fermi-liquid-like. It is still 
possible
that there is a plenty of phase structures at finite temperature.   
Because all $K_{\alpha\beta}$ are independent of each other, in general
there will be a unique smallest $K_{\alpha \beta}$ with $(\alpha\beta)=
(\alpha_0\beta_0)$. When $K_{\alpha_0\beta_0}$ decreases to 1 and then
smaller than 1, a phase transition happens.
There are such six phases that $g_{\alpha_0\beta_0}$
is relevant and all the $g_{\alpha\beta}\not=g_{\alpha_0\beta_0}$ are
irrelevant in each phase. They may have different finite temperature 
behaviors. For example, we consider $g_y$-relevant phase.
We focus on the fixed point Hamiltonian. Introducing the scaled parameters
$V_y/L=J_\perp/2$,
$V_{0\sigma}/L=-J_z/4$
and $V_{x\sigma}/L=J_z/2$ with $J_z>0$ (this ensures
$K_{\uparrow\downarrow}<1$),
we see the fixed point
Hamiltonian $H^*$ is just the one-channel Kondo Hamiltonian:
\begin{eqnarray}\displaystyle
H^*&=&\displaystyle\sum_{k>0,\sigma}
\epsilon_kc^\dagger_{k\sigma}c_{k\sigma}+\epsilon_d n_d
+Un_{d\uparrow}n_{d\downarrow}\\ \nonumber 
&+&\sum_{k>o,l=1}\frac{k}{\rho}b^\dagger_{kl}
b_{kl}
+\displaystyle\frac{J_\perp}{2}(S^+_c s^-_d+h.c.)+J_zS^z_cs^z_d.
\end{eqnarray}    
Here ${\bf S}_c$ and ${\bf s}_d$ are the spin operators of the hybridizing
and local electrons respectively. The $g_y$ ($J_\perp$)-relevant
phase is controlled by the strong fixed point of Kondo problem.
Furthermore, the values of $V_{0\sigma}$ and
$V_{x\sigma}$
are not important as long as they are within the parameter regime of
the $g_y$-relevant phase. Hence, the fixed point corresponding to
$g_y$-relevant phase is the FL, which is a collective mode of the system and
transfers to a state with a local moment at $T>T_K$, the Kondo temperature.
Equations (\ref{s332}) is reduced to
$$
\tau_0\frac{d\epsilon_d}{dln\tau_0}=
\tau_0\epsilon_d-g^2_y,~~~~~\tau_0\frac{dU}{dln\tau_0}=\tau_0 U+2g^2_y,
$$
which imply that $U$ increase positively if its initial value is not negative.
This repeats the results of Anderson model:
When the impurity level width $\Gamma\sim t^2_1$ is much less than $U$,
the model is equivalent to the Kondo model \cite{And3}.

For the $g_{1\sigma}$($\tilde g_{1\sigma}$)-relevant phase,
according to (\ref{s332}), $U$ keeps to be almost invariant if its initial 
value is the same as at the fixed point.
Hence, $\Gamma\gg U$ in this case and the groundstate of the system
may be read off from the exactly solvable Fano-Anderson model. It is
a FL but there is no a local moment in excitation states. So we call
this phase the free-orbital phase \cite{Wil1}.

For the $g_2$-relevant phase, 
the empty and doubly-occupied states favor over singly-occupied states.
The fixed point Hamiltonian $H^*=H^*_0+H^*_I$ where
$H^*_0$ can be read off from (\ref{s27}) with replacing the all of parameters 
by
their fixed point values and $H^*_I$ has a leading term $R
\sim (g^*_2/\tau_0)\sum_{\sigma}
(c^\dagger_\sigma c^\dagger_{-\sigma}d_\sigma d_{-\sigma}+h.c.)$.
The matrix elements of $R$ in the fixed point
can be evaluated by a similar considerations in the X-ray
edge problem \cite{32}. As done in \cite{Per1}, define $|n_d=0>$ to be the
eigenstates of $H^*_0|_{n_d=0}$ and $|n_d=2>$ to be $H^*|_{n_d=2}$'s.
Then
\begin{equation}
<n_d=0|R|n_d=2>\sim \tau^{\alpha}.
\label{R}
\end{equation}
The anomalous exponent $\alpha= (1-K_{03})/2$.
(If $\alpha<0$ the operator $R$ is irrelevant, and the fixed point is the NFL
as we have discussed.)
In the present parameter regime, $\alpha>0$ and
the operator $R$ are relevant. The fixed point is
regarded as the FL \cite{Per1}.

Thus, all phases with one relevant fugacity are FL but they may
have different finite temperature behaviors.
A FL-NFL phase transition occurs as the model parameters
vary.

(iii) When several of the $K_{\alpha\beta}\leq 1$, the
phase transition can be
characterized by Cardy's graphical method \cite{Car1}. Draw a graph whose 
vertices
are labeled by the states $0,\sigma$ and 3 (see Figure 1(a)).
A particular transition
corresponds to drawing a set of the edges $(\alpha\beta)$ which express that
those $K_{\alpha\beta}\leq 1$. When $K_{\alpha\beta}$ and $K_{\beta '\gamma}$
become smaller than one, $g_{\alpha\beta}$ and $g_{\beta'\gamma}$
are relevant.
If $\beta\not=\beta'$, $g_{\alpha\gamma}$ is still irrelevant
as long as $K_{\alpha\gamma}>1$. Before
the next phase transition happens, we change the edge linking the vertices
$\alpha$ and $\beta$ to a new vertex and the edge of $\beta'$ and $\gamma$
to another vertex. Then repeat the previous
process when other $K_{\alpha'\gamma'}
\leq 1$. If $\beta=\beta'$, even $K_{\alpha\gamma}>1$ the second
and third terms of (\ref{s31})
also drive $g_{\alpha\gamma}$ to be relevant. Reflecting this point
graphically, the vertices $\alpha,\beta$ and $\gamma$ are linked as an
edge which is changed to a new vertex before the next phase transition
appears. Cardy had explained the physical reason to cause $g_{\alpha\beta}$
becomes relevant \cite{Car1}: Logarithmic interactions between kinks are no 
longer
existent in the linked piece of the graph. Repeat the process till all
the vertices are linked so that no further transitions are possible.
Cardy calls
the final phase high temperature (HT)  phase, in which all the fugacities are
relevant. Notice that the renormalization group equations are not valid
quantitatively in the strongly coupling region. Therefore, the phase structure
of the model in the relevant phase must be discussed more carefully. In some
situations, the real phase diagram of the relevant phases
is different from that given by the graphical method. To see this we now 
turn to various reduced models from the general
model.

\subsection{ Spin Symmetric Model}

The spin symmetric model is a model with $g_{\alpha\uparrow}=g_{
\alpha\downarrow}$ and $K_{\alpha\uparrow}=K_{\alpha\downarrow}$. The phase
structure of the system can be described graphically in Figure 1.
Every diagram row of Figure 1 represents a sequence of transitions from
the NFL phase to HT phase. The diagram consisting only of the vertices
represents a phase while the diagram with
lines is a critical point.
For example, Figure 1(a) represents the sequence of transitions
corresponding to $K_{0\sigma}<K_{03},K_{3\sigma},K_{\uparrow\downarrow}$.
In fact, Fig. 1(a) and
1(b), 1(c) and 1(d) reflect similar physical processes, respectively. The
four-vertex phase in the figure is the NFL phase. The two-vertex phase
of 1(a)(1(b)) is the $g_{1\sigma}(\tilde g_{1,\sigma})$ and $g_y$ relevant
phase. We arrive at
the strong-coupling fixed point of the asymmetric Anderson model. This is
the FL. The three-vertex phase of 1(c) (1(d)) is controlled by the
strong-coupling fixed point of the Kondo problem. And two-vertex phase
is the same phase as 1(a)(1(b)).  The three-vertex phase of
1(e) is the $g_2$-relevant
phase, which is the FL. The phase structure of the spin symmetric model
is different from that of the general model: The free orbital phase does not
exist. Physically, since $g_{\alpha\uparrow}=g_{\alpha\downarrow}$
the spin can not be frozen out. Furthermore,
when the $(0\uparrow)$ kinks are bounded, the $(0\downarrow)$ kinks are also
bounded because of the spin symmetry, which leads to the $(\uparrow\downarrow)
$ kinks are bounded too. Therefore, $g_y$
becomes relevant together with $g_{1\sigma}$ or $\tilde g_{1\sigma}$.
The free orbital condition, $\Gamma\gg U$, can not be satisfied
in the spin symmetric model. We would like to mention a point that
the spin symmetry does not change the dimensions of the Lie algebra
described in Appendix A. Correspondingly, the largest number of the critical
points in the sequences of the phase transition keeps to be invariant.
As we have mentioned, in the strong coupling limit, the graphical method
is not always valid quantitatively. Here this point reflected in the
phase diagram is that the phase transitions from three-vertex phases to
two-vertex phases do not really happen because both of the asymmetric
Anderson model and the Kondo model have the same strongly coupling fixed point.

\subsection{\bf Particle-Hole Symmetric Models}

Setting $g_{1\sigma}=
\tilde g_{1,\sigma}$ and  $K_{0\sigma}=K_{-\sigma 3}$ in the general model,
the consistency of the renormalization flows of $g_{1\sigma}$ and $
\tilde g_{1\sigma}$
forces that $E_3=E_0$ (as long as their initial values are equal). 
Thus, if the PH
symmetry of the common hybridization
term in the model is not broken, the consistency of the renormalization
group equations requires that
the system keeps the symmetry. The phase
structure of the PH symmetric model is described in Figure 2. Since there is
no spin symmetry, the free orbital phase exists (the two-vertex phase
in Figs. 2(a) and 2(b)).
The phase described by two-vertex diagrams in Figure 1
disappears because the model is PH symmetric. The Kondo phase and
$g_2$-relevant phase still exist.

If we consider spin and PH symmetric model,
the phase structure is shown by Figures 3(a),(b) and (c). This model has been
discussed in \cite{Yu1}. Figures 3(a), (b) and (c) can be
reexpressed more explicitly by Figure 3(d).
Notice that there is no three critical point sequence of the transition in the
PH symmetric model because it is impossible that $K_{\uparrow\downarrow}<
K_{03}<K_{0\sigma}$ and $K_{03}<K_{\uparrow\downarrow}<K_{0\sigma}$.
Physically, those sequences of transition break the PH symmetry and
then are forbidden. Relating to the Lie algebra of the plasma `charges',
under the PH symmetry, the dimensions of the Lie algebra reduce to 3.
As we have done \cite{Yu1}, the strongly coupling limit needs to be treated
more carefully. The real phase diagram is shown in Figure 3(e).
The phase of all fugacity relevance is divided into three
parts: $I$ is still in the Kondo strongly coupling phase; $II$ is the same as
$G_2$-phase in which the empty and doubly occupied states are favored;
and $III$ is all impurity state mixed phase.

\section{Mixed Valence state regime and fixed points}

In the previous section, we term the FL and NFL states only in the sense that
whether the infrared orthogonality catastrophe of the X-ray edge type 
appears. In fact, the situation may be subtler if the we consider the
local state properties in the problem \cite{Si1,SK}. To see this, we study 
the fixed points in the mixed valence state regimes, in which the
renormalized impurity level shift $\epsilon_d^*$ is of the same order as  
the renormalized resonant width $\Gamma^*=\pi g^{*2}_t/\tau$, i. e.,
$|\epsilon_d^*|\leq \Gamma^*$. Most interesting 
physical phenomena appear in the particle-hole symmetric model. At finite 
$U$, the particle-hole symmetry implies $\epsilon_d=-U/2$ for their bare 
values. The renormalization group equations show that this constraint
is kept after the renormalization except in the case of infinite $U$. So,
the crossover between various regimes is still determined by the 
ratio  $|\epsilon_d^*|/\Gamma^*$. We can still define the mixed valence 
states if $|\epsilon_d^*|\sim \Gamma^*$. On the other hand, the 
particle-hole symmetry reduces the number of the independent stiffness constants 
to two 
, $\gamma_0$ and $\gamma_x$. The fixed points in the $\gamma_0-\gamma_x$
plane can be figured out according to the stiffness constants
$\epsilon_{t_1}=\frac{1}{2}\gamma_0$,  $\epsilon_x=\gamma_x$
$\epsilon_{t_2}=2\gamma_0-\gamma_x$. The weak coupling mixed valence states
is within the renormalized parameter regime $\epsilon^*_{t_1}>1$
, $\epsilon^*_x>1$ and $\epsilon^*_{t_2}>1$. The strong coupling mixed
valence states are corresponding to  $\epsilon^*_{t_1}<1$ and 
$\epsilon^*_{t_2}<1$. There are two
intermediate coupling mixed valence state regimes in which the renormalized 
stiffness constants are given by 1) $\epsilon^*_{t_1}<1$,  $\epsilon^*_x<1$
and  $\epsilon^*_{t_2}>1$ and 2)  $\epsilon^*_{t_1}>1$,  $\epsilon^*_x<1$
and  $\epsilon^*_{t_2}<1$. The latter associates to the intermediate 
coupling mixed valence state we have already known 
in the infinite $U$ case \cite{Hald3,SK} while the former 
to the $G_2$-irrelevant mixed valence states
we newly find in the finite $U$ case here. We will not repeat the fixed 
point behaviors in the strong coupling case and the known intermediate 
coupling case but refer to the interested readers to read the relevant 
literatures \cite{Hald3,SK}. What we would like to do is to review the
weak coupling states and focus on the $G_2$-irrelevant states.

Within the weak coupling mixed valence regime, all running fugacities could
be obtained by solving the linear renormalization group equations. As we have 
seen in the last 
section, there is a NFL fixed point in the sense of the X-ray edge
infrared orthogonality catastrophe. Furthermore, in the mixed valence regime, 
there are also
the violation of the Fermi liquid behavior in the calculation of the 
local single-particle and two-particle correlation functions. Some of them
have been shown in refs.\cite{Si1,SK} such as $G_{dc}, G_{dd}$ and
two-particle excitonic and $d$-electron correlation functions as well as
$d c$ correlation functions. Divergences appear in the 
$G_{dc}$, $G_{dd}$ and the $d c$ correlation functions, which shows the 
local NFL behavior. At the finite $U$, one may have more divergent
correlation functions. For instance, the single $d$-electron paring
function
\begin{eqnarray}
&&G_{\rm d-pair}(\tau)=<0|d_\sigma(\tau) d_{\sigma'}(0)|3>\sim \tau
^{-\alpha_d},\nonumber\\
&&{\rm or} \nonumber\\
&&G_{\rm d-pair}(\omega)\sim \omega^{-1+\alpha_d},
\end{eqnarray}
with the exponent 
$$\alpha_d=2\delta_0^{*2}-\delta_x^{*2}(1-\delta_{\sigma,\sigma'}).$$
The two-particle $d$-electron paring correlation function is given rise to
\begin{equation}
<d_\sigma d_{\sigma'}(\tau ) d_{\sigma'}^\dagger d_{\sigma}^\dagger(0)>
\sim \tau^{-\alpha_4},
\end{equation}
with the exponent
$$
\alpha_4=-\biggl(\frac{\delta^*_0+\delta_x^{*2}}{2}\biggr)^2.
$$

We see that both $d$-electron pairing correlation functions are divergent 
in the low energy limit.

The $G_2$-irrelevant intermediate coupling mixed valence state is newly
discovered at the finite $U$ model. The renormalization group 
equations yield an initially
decreasing the paring charge fugacity $g_2$ and initially increasing the
charge and spin fugacities $g_1$ and $g_x$. It can occur for which the 
spin and charge kinks are unbound while the paring charge kinks are bound.
The two-body hybridization is irrelevant while the hybridization and exchange
coupling are relevant. Therefore, the correlation functions
such as $<0|R|3>$ given by (\ref{R}) and local pairing single- and 
two-$d$-electron correlation functions are still divergent. This gives a new 
set of local NFL fixed points.

\section{Conclusions}

We have constructed a general impurity model with two-body hybridization.
Through comparing with the  partition functions, via bosonization, we have found
that this impurity problem is asymptotically
equivalent to a six-component plasma of
kinks with external fields. Hence, the renormalization group analysis
can be systematically carried out in the framework of the AYHC poor man's
scaling theory.
We emphasize that it is important to introduce the two-body
hybridization in the Hamiltonian. This ensures the transitivity of the
`dipole' operators and then the equivalence of the partition functions.
Moreover, the renormalization group equations show that the flows excluding
the two-body hybridization are not correct renormalization flows. In
some phase of the phase
diagram, $g_2$ is relevant, {\it i.e.}, hybridizing electrons hop pairly onto
and off the impurity orbital . This behavior of the electrons
resembles the `Cooper pair' in BCS theory. 
The property of the $g_2$-relevant phase is not clear yet. Perhaps, it relates
to the superconduct phase in an impurity lattice model. 
In the phase diagram, there is a NFL phase which
controlled by the fixed point behavior of the multi-channel X-ray
edge problem. Around this phase we have six other phases called
the $g_2$-relevant phase, the strong-coupling Kondo phase and four equivalent
free orbital metallic phases.
Hence, there is a FL-NFL phase transition in the system. There are six
sequences of phase transitions from the NFL phase to the HT phase in which
all fugacities are relevant. Each sequence has three critical points and
four phases, which
is same as the dimensions of the Lie algebra corresponding to the plasma
`charges'.

We have discussed various reduced models. For the spin symmetric model,
the free orbital states do not exist because we can not freeze only
one spin state in the impurity orbital and the free orbital condition
can not be satisfied. For the PH symmetric model, we addressed that
the PH symmetry of the common one-body hybridization term in the Hamiltonian
forces $\epsilon_d=-U/2$ by the consistency of the renormalization group
equations and also decrease a critical point in the sequence of transitions,
correspondingly reducing a dimension of the `charges' Lie algebra.

We also discussed the fixed point behaviors in the mixed valence state regime.
Except the X-ray edge type infrared orthogonality catastrophe in the screening 
channel, 
we also found the local NFL signal in the weak coupling and intermediate 
coupling mixed valence states.
\bigskip

\noindent{\bf Acknowledgments}

One of us (Y. Y.) is grateful to Y. M. Li and N. d'Amberumenil for their
co-operation in the earlier stage of this work and useful discussions. The 
authors thank Z. B. Su for discussions. This work was supported in part
by NSF of China and Grant LWTZ-1298 of Chinese Academy of Science.

\appendix

\section{Lie Algebra and Its Relation to Phase Structures}

\bigskip

In this appendix, we discuss some mathematical properties of the `charges'
$q^a_{\alpha\beta}$. The screening components, $q^l_{\alpha\beta}$, are
trivial and we focus on the spin-components, $q^\sigma_{\alpha\beta}$.
We define $q^\gamma_{\alpha\beta}$  are given by the spin-components
for $\gamma=\sigma$ or vanish for $\gamma=0$ and $3$.
In terms of the transitivity (\ref{s222}) and the anti-symmetry
of the `charges', we immediately have $q^\gamma_{\alpha\beta}$
obey the Jacobi identities:
\begin{equation}
\sum_\gamma q^\gamma_{\alpha\beta}q^\lambda_{\gamma\delta}+q^\gamma_{\beta
\delta}q^\lambda_{\gamma\alpha}+q^\gamma_{\delta\alpha}q^\lambda_{\gamma
\beta}=0, 
\label{a1}
\end{equation}
which means that the `charges' may be regarded as a set of structure constants
of some Lie algebra. Denoting the generators of
the Lie algebra by $T_\alpha$, then we have
\begin{eqnarray}
[T_0,T_\sigma]=q^{\sigma'}_{0\sigma}T_{\sigma'},~~~~[T_\sigma,T_{\sigma'}]=q^{
\sigma''}_{\sigma\sigma'}T_{\sigma''},\label{a2}\\ \nonumber
[T_\sigma,T_3]=q^{\sigma'}_{\sigma 
3}T_{\sigma'},~~~~[T_0,T_3]=q^\sigma_{03}T_\sigma.
\end{eqnarray}
The Lie algebra has the following properties:

(i) It is solvable. We call the Lie algebra ${\cal L}$. Equations (\ref{a2})
imply that $[{\cal L},{\cal L}]\subseteq {\cal L}_1$, $[{\cal L}_1,
{\cal L}_1]\subseteq
{\cal L}_2$ and $[{\cal L}_2,{\cal L}_2]=0$, where ${\cal L}_1=\{T_\sigma\}$
and ${\cal L}_2=\{q_{\uparrow\downarrow}^\sigma T_\sigma\}$.
This means that ${\cal L}$ is solvable.

(ii) The Lie algebra is 4-dimensional, which is
same as the dimensions of the impurity Hilbert space.
The structure constants form
the adjoint representation by $(T_\alpha)_{\beta\gamma}=q^\gamma_{\alpha
\beta}$. For the  general parameters $\delta_0,\tilde\delta_0,
\delta_{x\sigma}$ and $\tilde\delta_{x\sigma}$, $aT_0+bT_\uparrow
+c T_\downarrow
+dT_3=0$ only if all coefficients $a,b,c$ and $d$ are zero.

(iii) In the infinite $U$ limit, the doubly-occupied state is completely
suppressed in the discussion of the low energy behavior of the system.
The physics has been discussed in \cite{Per1}\cite{Si1,SK}. The impurity Hilbert 
space 
is
3-dimensional, and correspondingly, the Lie algebra reduces to
3-dimensional one: ${\cal L}=\{T_0,T_\sigma\}$.

(iv) In the PH symmetric model, $q^\gamma_{0\sigma}=q^\gamma_{-\sigma,3
}$, which leads to the dimensions of the Lie algebra is reduced to 3 too.
This consequence can be checked in the adjoint representation by the
following linear transformation
\begin{eqnarray}
L_0=T_0+T_3+T_\uparrow+T_\downarrow,\\\nonumber
L_1=T_0-T_3+T_\uparrow+T_\downarrow,\\\nonumber
L_2=T_0-T_3+T_\uparrow-T_\downarrow,\\ \nonumber
L_3=T_0+T_3-T_\uparrow-T_\downarrow.
\end{eqnarray}
The transformation is a non-singular but one finds that $L_3\equiv 0$. This
implies that the dimensions of the Lie algebra is 3.

At the moment, it is not clear that what are the physical implication
of the Lie algebra and the corresponding Lie group. We only know that there is
a correspondence between the phase structures and the dimensions, $D$,
of the Lie
algebra: $D-1$ is equal to the largest numbers of the critical lines
between the NFL phase and the HT phase.

\section{  Renormalization Group Equations}

Our partition function (\ref{s221}) has arrived at a special form of the Cardy's
general model with an external field added. Therefore, the renormalization
group equations can be directly derived by using the AYHC poor man's
scaling theory. The details of the derivation have been presented in 
\cite{And4}\cite{Car1}\cite{Si1,SK}. Here we just address something in their 
derivation.

The renormalization group equations describe the flows of the dimensionless
couplings, the fugacities $g_{\alpha\beta}$, the stiffness constants $K(
\alpha,\beta)$, and the external fields $E_\alpha$, as the band width
is reduced. 

Firstly, when the cut-off $\tau_0 \rightarrow \tau_0+d\tau_0$,
to compensate the changes of $\tau_0$'s power terms and external 
field terms in the partition function (\ref{s221}),
the fugacities and the external
fields change as follows:
\begin{eqnarray}
\delta g_{\alpha_{i+1}\alpha_i}&=&(dln\tau_0)
g_{\alpha_{i+1}\alpha_i}(1+K(\alpha_{i+1},\alpha_i)),\\ \nonumber
\delta E_{\alpha_i}&=&E_{\alpha_i}dln\tau_0.
\end{eqnarray}

The cut-off of integral limits also causes the variation
of the logarithmic terms. After integrating out the $\tau_{i+1}$,
we obtain the variation involving two such kinks
$i, i+1$ and a third kink $j$ which is given by
\begin{eqnarray}
\displaystyle &&(dln\tau_0) g_{\alpha_i\alpha_{i+1}}g_{\alpha_{i+1}\alpha_{i+2}}
\\ \nonumber
&&\exp\{
\tilde K(\alpha_i,\alpha_{i+1};\alpha_j,\alpha_{j+1})ln\frac{\tau_j-\tau_i}{
\tau_0}\\ \nonumber
&&+\tilde K(\alpha_{i+1},\alpha_{i+2};\alpha_j,\alpha_{j+1})ln
\frac{\tau_j-\tau_i-\tau_0}{\tau_0}\}\\ \nonumber
&&\exp\{E_{\alpha_{i+2}}-E_{\alpha_{i+1}}\}
\\ \nonumber \displaystyle
&&\approx (dln\tau_0) g_{\alpha_i\alpha_{i+1}}g_{\alpha_{i+1}\alpha_{i+2}}
\\ \nonumber &&\exp\{(K(\alpha_i,\alpha_j)+K(\alpha_{i+2},
\alpha_{j+1})\\ \nonumber
&&-K(\alpha_i,
\alpha_{j+1})-K(\alpha_{i+2},\alpha_j))ln\frac{\tau_j-\tau_i}{\tau_0}\}
\\ \nonumber &&\exp{E_{\alpha_{i+2}}-E_{\alpha_{i+1}}}.
\end{eqnarray}
The only approximation is $\tau_j-\tau_i \gg \tau_0$, which means the kinks
are rare enough.

As long as $\alpha_i\not=\alpha_{i+2}$, this can be incorporated into
a renormalization of $g_{\alpha_i\alpha_{i+1}}$, {\it i.e.}
\begin{eqnarray}
\delta g_{\alpha_i\alpha_{i+1}}&=&
(dln\tau_0)\sum_\alpha g_{\alpha_i\alpha}g_
{\alpha\alpha_{i+1}}e^{E_{\alpha_{i+1}}-E_\alpha},
\end{eqnarray}
which together with (B.1) and $g_{\alpha\beta}=g_{\beta\alpha}$ leads to the 
renormalization flows of the fugacities
in arbitrary external fields:
\begin{equation}
\frac{dg_{\alpha\beta}}{dln\tau_0
}=(1+K(\alpha,\beta))g_{\alpha\beta}+
\sum_\gamma g_{\alpha\gamma}g_{\gamma\beta}e^{-E_\gamma+\frac{1}{2}
(E_\alpha+E_\beta)}.
\label{b3}
\end{equation}

However, for the neutral pair with $\alpha_i=\alpha_{i+2}$, the leading term
of the left hand side of (B.2) is independent of the relative position of
the kinks. It becomes:
$$
(dln\tau_0)g^2_{\alpha_i\alpha_{i+1}}\frac{\tau_{i+2}-\tau_{i-1}}{\tau_0}
e^{E_{\alpha_i}-E_{\alpha_{i+1}}}.
$$
which leads to the renormalization of the external fields:
$$
\delta E_\alpha=-dln\tau_0\sum_\beta g^2_{\alpha\beta}e^{E_\alpha-E_\beta}.
$$

Taking into account $\sum E_{\alpha}=0$ and the renormalization of the 
free energy $F$ , the
renormalization flows of $E_\alpha$ and $F$ are
\begin{eqnarray}
&&\displaystyle\frac{dE_\alpha}{dln\tau_0}=E_\alpha-\sum_\gamma 
g^2_{\alpha\gamma}
e^{E_\alpha-E_\gamma}+\frac{1}{4}\sum_{\beta\gamma}g^2_{\beta\gamma}
e^{E_\beta-E_\gamma}, \label{b4}\\
[3mm]
\displaystyle 
&&\frac{dF\tau_0}{dln\tau_0}=-\frac{1}{4}\sum_{\alpha\beta}g^2_{\alpha
\beta}e^{E_\alpha-E_\beta}.
\label{b5}
\end{eqnarray}
Then, we must consider the next term in the expansion of (B.2) as
$\tau_0/(\tau_j-\tau_i)$, which is
\begin{eqnarray}                     \displaystyle
&&(dln\tau_0) g^2_{\alpha_i\alpha_{i+1}}\tilde K(\alpha_i,\alpha_{i+1};\alpha_j,
\alpha_{j+1})(ln(\tau_j-\tau_{i+2}) \\ \nonumber
&&-ln(\tau_j-\tau_{i-1}))
e^{E_{\alpha_i}-E_{\alpha_{i+1}}},
\end{eqnarray}
after integrated over the `closed pair' \cite{And4} range
$\tau_{i-1}\leq \tau_i\leq\tau_{i+2}$. This can be interpreted as the
renormalization of the stiffness constants:
\begin{eqnarray}      \displaystyle
\displaystyle\frac{dK(\alpha,\beta)}{dln\tau_0}&=&
\displaystyle-\sum_\gamma g^2_{\alpha\gamma}
e^{E_\alpha-E_\gamma} \label{b6}\\ \nonumber
&&(K(\alpha,\beta)+K(\alpha,\gamma)-K(\beta,\gamma))\\ \nonumber
&-&\displaystyle\sum_\gamma
g^2_{\beta\gamma}
e^{E_\beta-E_\gamma}\\ \nonumber
&&(K(\alpha,\beta)+K(\beta,\gamma)-K(\alpha,\gamma)).
\end{eqnarray}

Equations (\ref{b3})-(\ref{b5}) and (\ref{b6}) are our renormalization 
group equations in
the model. The relations $g_{\alpha, \alpha}=0$ and $K(\alpha,\alpha)=0$
and $\sum E_\alpha=0$ are
preserved in the renormalization process.

\eject

\centerline{Figure Captions}

\begin{description}
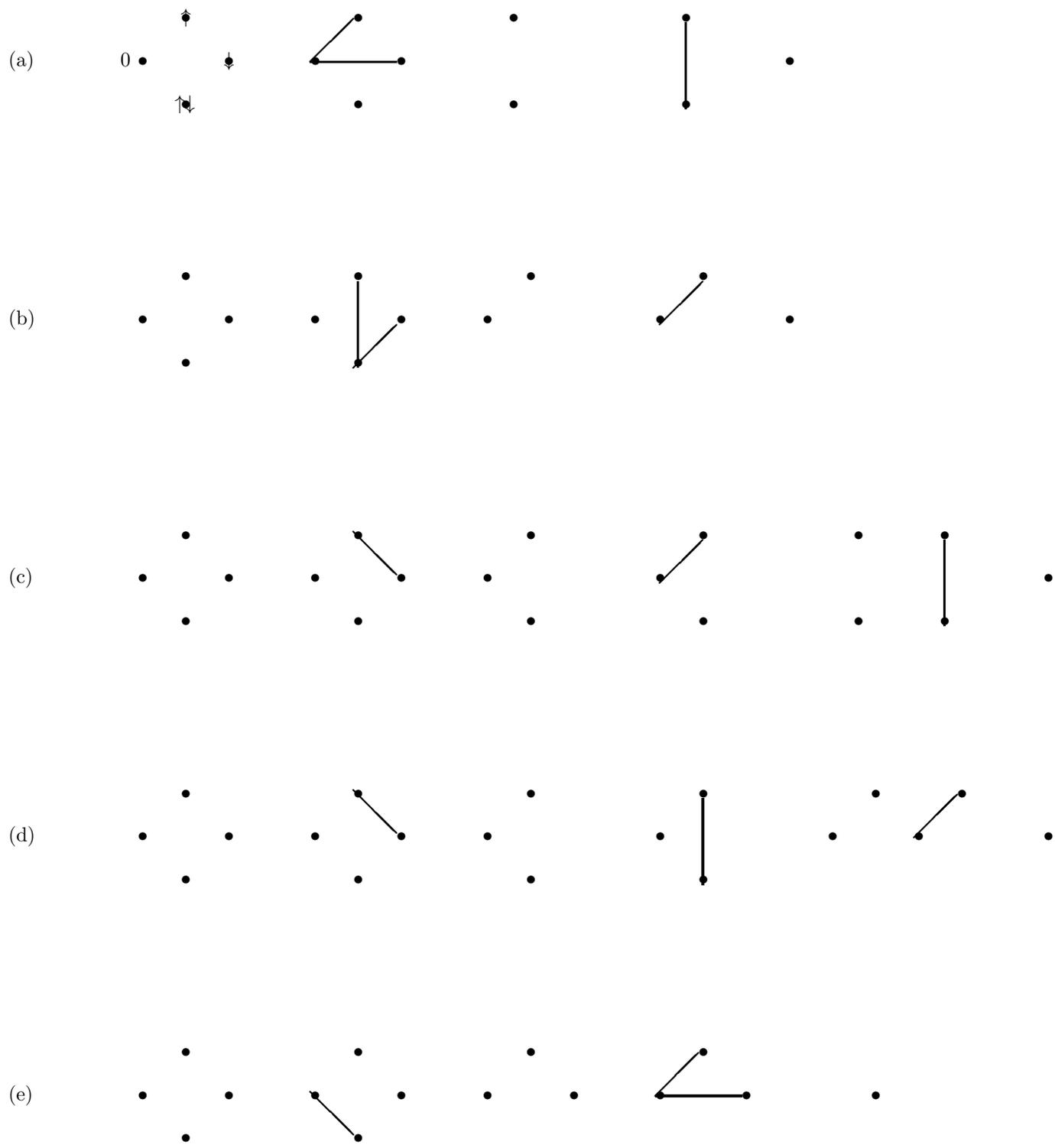
\item{Fig. 1.} The phase structure of the spin symmetric model. (a)
$K_{0\sigma}<K_{\sigma 3},K_{03}, K_{\uparrow\downarrow}$.
(b) $K_{\sigma 3}<K_{0\sigma},K_{03},K_{\uparrow\downarrow}$.
(c) $K_{\uparrow\downarrow}<K_{0\sigma}<K_{\sigma 3},K_{03}$.(d)
$K_{\uparrow\downarrow}<K_{\sigma}<K_{0\sigma},K_{03}$. (e) $K_{03}
<K_{0\sigma},K_{\sigma 3},K_{\uparrow\downarrow}$.
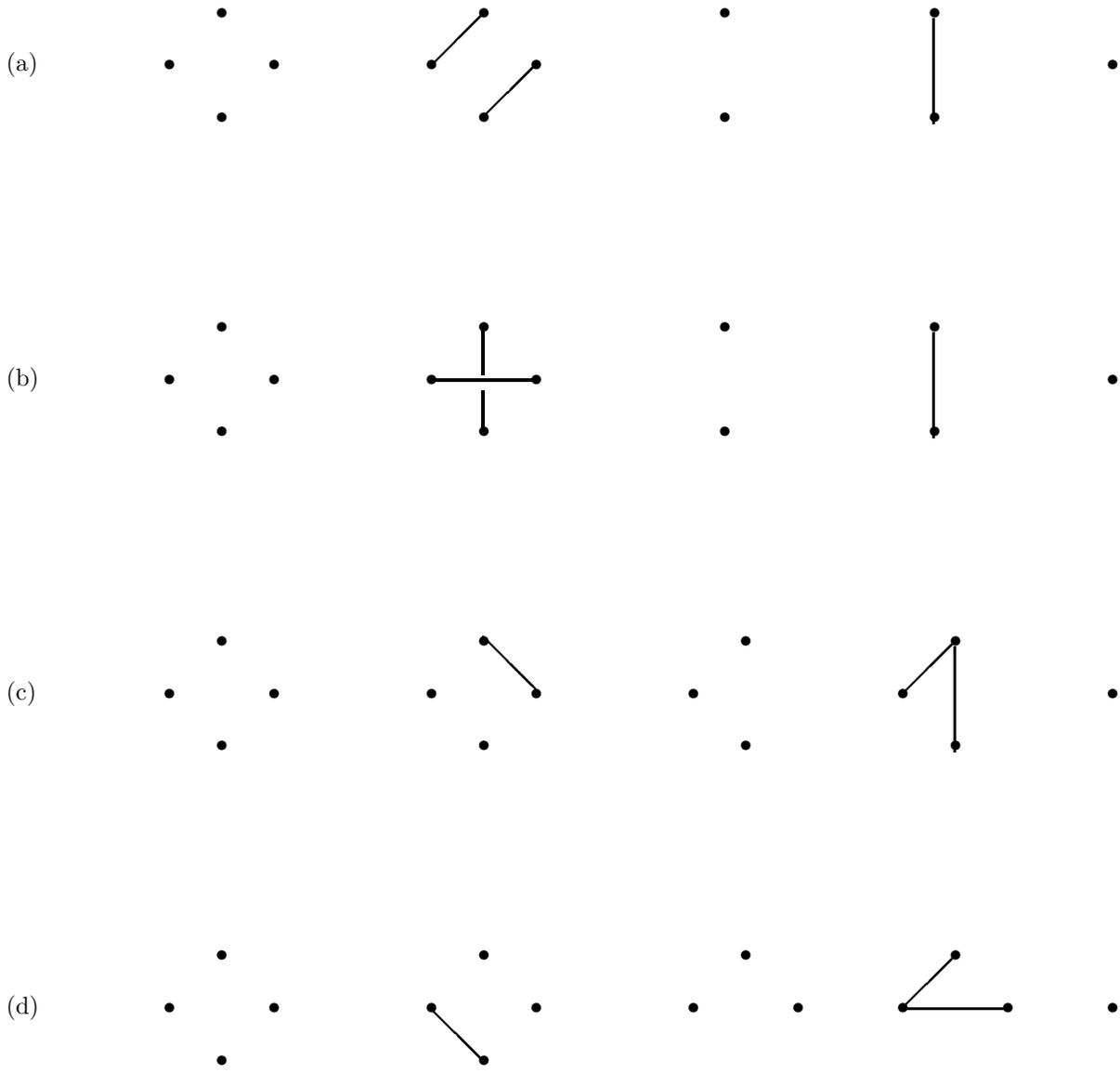
\item{Fig. 2.} The phase structure of the PH symmetric model. (a)
$K_{0\uparrow}<K_{0\downarrow},K_{\uparrow\downarrow},K_{03}$;
(b) $K_{0\downarrow}<K_{0\uparrow},K_{\uparrow\downarrow},K_{03}$;
(c)$ K_{\uparrow\downarrow}<K_{0\sigma}<K_{03}$; (d) $K_{03}<
K_{0\sigma}<K_{\uparrow\downarrow}$ or $K_{03}<K_{0\sigma}<K_{\uparrow
\downarrow}<K_{0,-\sigma}$.
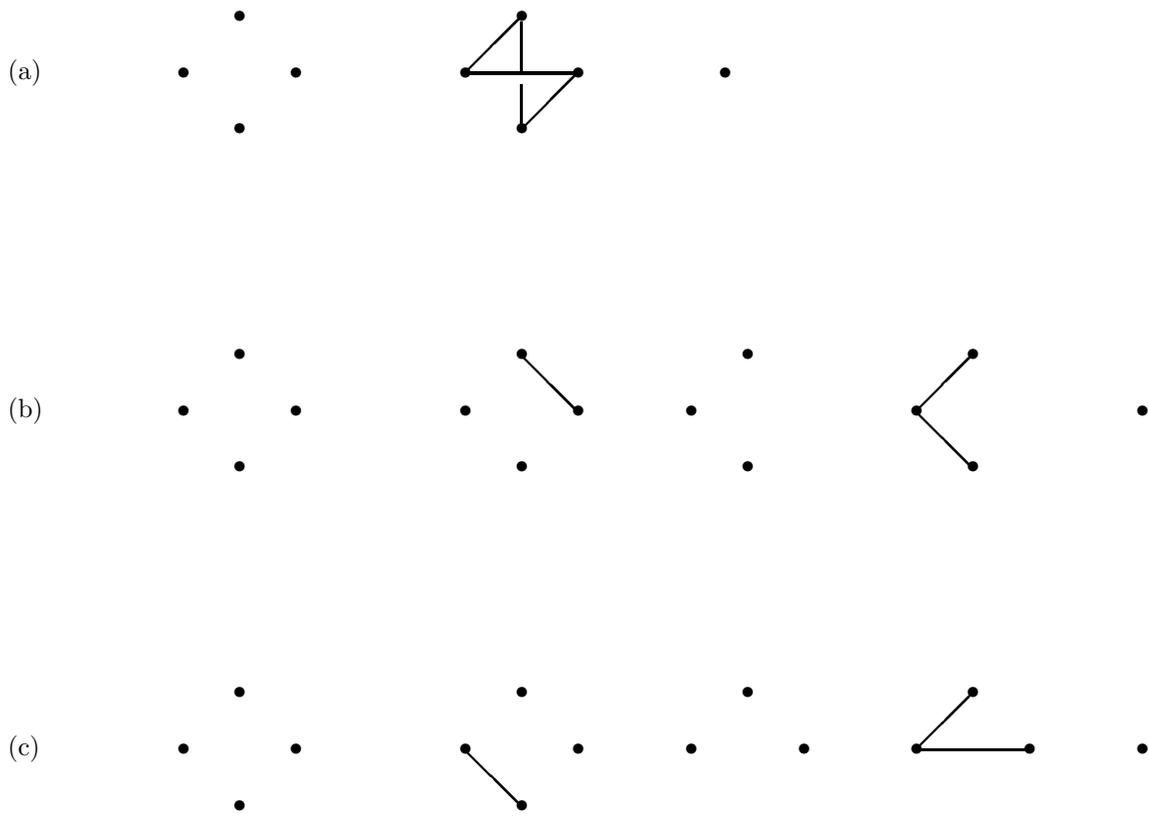
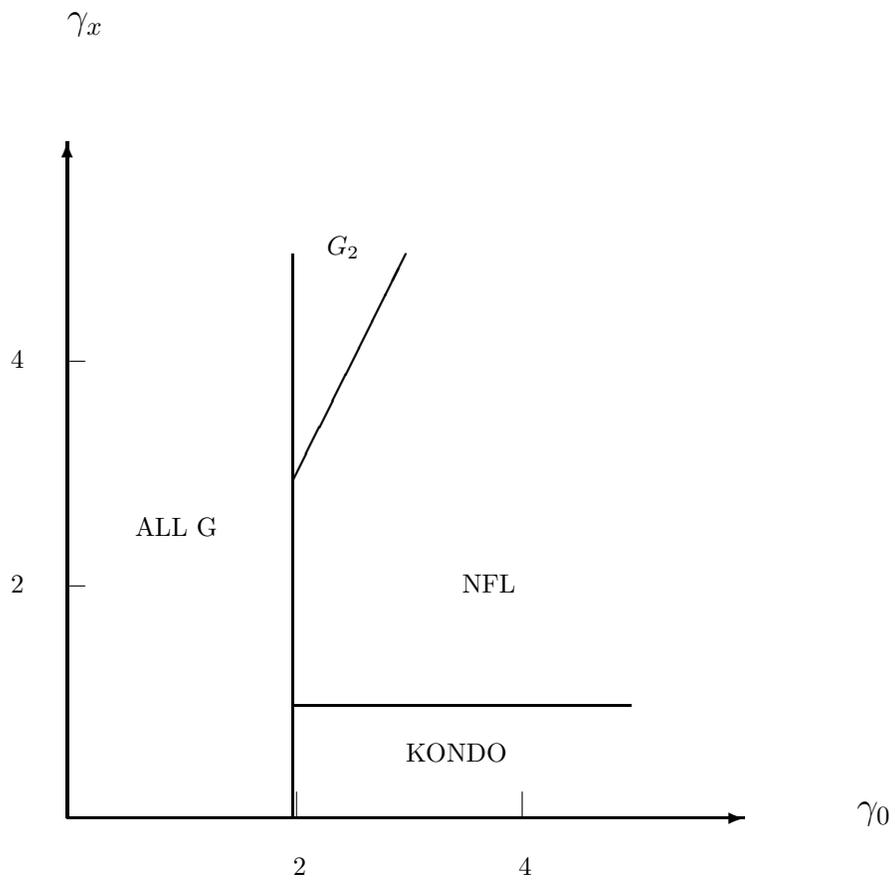
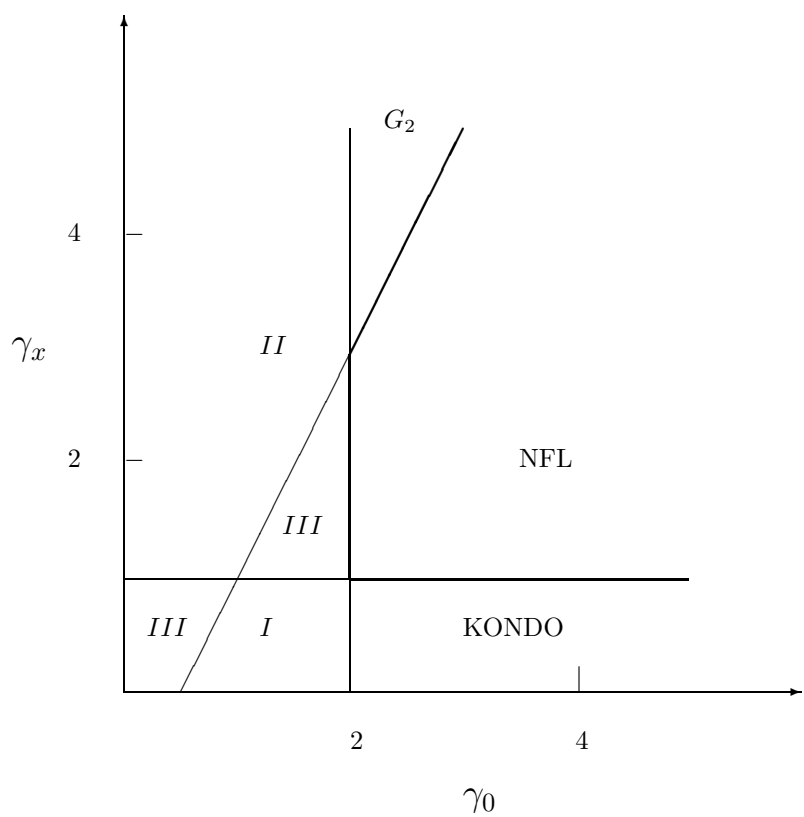
\item{Fig. 3.} The phase structure of the
spin and PH symmetric model. (a) $K_{0\sigma}=K_{\uparrow\downarrow}= K_{03}$;
(b)$ K_{\uparrow\downarrow}<K_{0\sigma}<K_{03}$; (c) $K_{03}<K_{0\sigma}<
K_{\uparrow\downarrow}$. (d) The phase diagram
in $\gamma_0-\gamma_x$ space, where
$\gamma_0=(1-\delta_x/\pi-\delta_0/\pi)^2+(\delta_0/\pi)^2+\sum_l(\delta_l/\pi
)^2$ and $\gamma_x=(1-\delta_x/\pi)^2$.
(e) The real phase diagram in $\gamma_0-\gamma_x$ space. 
The thick lines are phase boundaries.
The thin lines divide the Fermi liquid phase into the different regions
characterized by different behaviors at finite temperature.
\end{description}

\end{multicols}

\newpage

\begin{picture}(8,8)(-10,550)
\unitlength 1.5cm
\thicklines
\put(0.0,0.0){\shortstack{$\bullet$}}
\put(.5,.5){\shortstack{$\bullet$}}
\put(.5,-.5){\shortstack{$\bullet$}}
\put(1.0,0.0){\shortstack{$\bullet$}}
\put(.0,3.0){\shortstack{$\bullet$}}
\put(.5,3.5){\shortstack{$\bullet$}}
\put(.5,2.5){\shortstack{$\bullet$}}
\put(1.0,3.0){\shortstack{$\bullet$}}
\put(.0,6.0){\shortstack{$\bullet$}}
\put(.5,6.5){\shortstack{$\bullet$}}
\put(.5,5.5){\shortstack{$\bullet$}}
\put(1.0,6.0){\shortstack{$\bullet$}}
\put(.0,9.0){\shortstack{$\bullet$}}
\put(.5,9.5){\shortstack{$\bullet$}}
\put(.5,8.5){\shortstack{$\bullet$}}
\put(1.0,9.0){\shortstack{$\bullet$}}
\put(.0,12.0){\shortstack{$\bullet$}}
\put(.5,12.5){\shortstack{$\bullet$}}
\put(.5,11.5){\shortstack{$\bullet$}}
\put(1.0,12.0){\shortstack{$\bullet$}}
\put(-.2,12.0){\shortstack{0}}
\put(.5,12.5){\shortstack{$\uparrow$}}
\put(1.0,12.0){\shortstack{$\downarrow$}}
\put(.43,11.5){\shortstack{$\uparrow\downarrow$}}
\put(2.0,0.0){\shortstack{$\bullet$}}
\put(2.5,.5){\shortstack{$\bullet$}}
\put(2.5,-.5){\shortstack{$\bullet$}}
\put(3.0,.0){\shortstack{$\bullet$}}
\put(2.0,.1){\line(1,-1){0.5}}
\put(2.0,3.0){\shortstack{$\bullet$}}
\put(2.5,3.5){\shortstack{$\bullet$}}
\put(2.5,2.5){\shortstack{$\bullet$}}
\put(3.0,3.0){\shortstack{$\bullet$}}
\put(2.5,3.6){\line(1,-1){0.5}}
\put(2.0,6.0){\shortstack{$\bullet$}}
\put(2.5,6.5){\shortstack{$\bullet$}}
\put(2.5,5.5){\shortstack{$\bullet$}}
\put(3.0,6.0){\shortstack{$\bullet$}}
\put(2.5,6.6){\line(1,-1){.5}}
\put(2.0,9.0){\shortstack{$\bullet$}}
\put(2.5,9.5){\shortstack{$\bullet$}}
\put(2.5,8.5){\shortstack{$\bullet$}}
\put(3.0,9.0){\shortstack{$\bullet$}}
\put(2.5,8.5){\line(1,1){.5}}
\put(2.55,8.5){\line(0,1){1.0}}
\put(2.0,12.0){\shortstack{$\bullet$}}
\put(2.5,12.5){\shortstack{$\bullet$}}
\put(2.5,11.5){\shortstack{$\bullet$}}
\put(3.0,12.0){\shortstack{$\bullet$}}
\put(2.0,12.05){\line(1,1){.5}}
\put(2.0,12.05){\line(1,0){1.0}}
\put(4.0,0.0){\shortstack{$\bullet$}}
\put(4.5,.5){\shortstack{$\bullet$}}
\put(5.0,0.0){\shortstack{$\bullet$}}
\put(6.0,0.0){\shortstack{$\bullet$}}
\put(6.5,.5){\shortstack{$\bullet$}}
\put(7.0,0.0){\shortstack{$\bullet$}}
\put(6.0,0.05){\line(1,1){.5}}
\put(6.0,0.05){\line(1,0){1.0}}
\put(8.5,0.0){\shortstack{$\bullet$}}
\put(4.0,3.0){\shortstack{$\bullet$}}
\put(4.5,3.5){\shortstack{$\bullet$}}
\put(4.5,2.5){\shortstack{$\bullet$}}
\put(6.0,3.0){\shortstack{$\bullet$}}
\put(6.5,3.5){\shortstack{$\bullet$}}
\put(6.5,2.5){\shortstack{$\bullet$}}
\put(6.55,2.5){\line(0,1){1.0}}
\put(8.5,3.5){\shortstack{$\bullet$}}
\put(8.0,3.0){\shortstack{$\bullet$}}
\put(9.00,3.0){\shortstack{$\bullet$}}
\put(9.5,3.5){\shortstack{$\bullet$}}
\put(9.00,3.05){\line(1,1){.5}}
\put(10.5,3.0){\shortstack{$\bullet$}}
\put(6.0,6.0){\shortstack{$\bullet$}}
\put(6.5,6.5){\shortstack{$\bullet$}}
\put(6.5,5.5){\shortstack{$\bullet$}}
\put(6.05,6.0){\line(1,1){.5}}
\put(4.0,6.0){\shortstack{$\bullet$}}
\put(4.5,6.5){\shortstack{$\bullet$}}
\put(4.5,5.5){\shortstack{$\bullet$}}
\put(8.3,6.5){\shortstack{$\bullet$}}
\put(8.3,5.5){\shortstack{$\bullet$}}
\put(9.3,6.5){\shortstack{$\bullet$}}
\put(9.3,5.5){\shortstack{$\bullet$}}
\put(9.35,5.5){\line(0,1){1.0}}
\put(10.5,6.0){\shortstack{$\bullet$}}
\put(4.0,9.0){\shortstack{$\bullet$}}
\put(4.5,9.5){\shortstack{$\bullet$}}
\put(6.0,9.0){\shortstack{$\bullet$}}
\put(6.5,9.5){\shortstack{$\bullet$}}
\put(6.05,9.0){\line(1,1){.5}}
\put(7.5,9.0){\shortstack{$\bullet$}}
\put(4.3,12.5){\shortstack{$\bullet$}}
\put(4.3,11.5){\shortstack{$\bullet$}}
\put(6.3,12.5){\shortstack{$\bullet$}}
\put(6.3,11.5){\shortstack{$\bullet$}}
\put(6.35,11.5){\line(0,1){1.0}}
\put(7.5,12.0){\shortstack{$\bullet$}}
\put(4.0,-2.0){\shortstack{FIGURE 1}}
\put(-1.5,0.0){\shortstack{(e)}}
\put(-1.5,3.0){\shortstack{(d)}}
\put(-1.5,6.0){\shortstack{(c)}}
\put(-1.5,9.0){\shortstack{(b)}}
\put(-1.5,12.0){\shortstack{(a)}}

\end{picture}
\eject
\begin{picture}(8,8)(-30,450)
\unitlength 1.5cm
\thicklines
\multiput(.0,.0)(0.0,3.0){4}{\shortstack{$\bullet$}}
\multiput(.5,-.5)(0.0,3.0){4}{\shortstack{$\bullet$}}
\multiput(.5,.5)(0.0,3.0){4}{\shortstack{$\bullet$}}
\multiput(1.0,0.0)(0.0,3.0){4}{\shortstack{$\bullet$}}
\multiput(2.5,0.0)(0.0,3.0){4}{\shortstack{$\bullet$}}
\multiput(3.0,-.5)(0.0,3.0){4}{\shortstack{$\bullet$}}
\multiput(3.0,.5)(.0,3.0){4}{\shortstack{$\bullet$}}
\multiput(3.5,0.0)(.0,3.0){4}{\shortstack{$\bullet$}}
\put(2.55,0.05){\line(1,-1){.5}}
\put(3.05,3.6){\line(1,-1){.5}}
\put(2.55,6.05){\line(1,0){1.0}}
\put(3.05,5.55){\line(0,1){0.4}}
\put(3.05,6.1){\line(0,1){.45}}
\put(2.55,9.05){\line(1,1){0.5}}
\put(3.05,8.56){\line(1,1){.5}}
\multiput(5.0,.0)(0.0,3.0){2}{\shortstack{$\bullet$}}
\put(6.0,.0){\shortstack{$\bullet$}}
\put(5.5,2.5){\shortstack{$\bullet$}}
\multiput(5.5,0.5)(.0,3.0){2}{\shortstack{$\bullet$}}
\multiput(5.3,5.5)(.0,3.0){2}{\shortstack{$\bullet$}}
\multiput(5.3,6.5)(.0,3.0){2}{\shortstack{$\bullet$}}
\multiput(7.0,.0)(.0,3.0){2}{\shortstack{$\bullet$}}
\put(8.0,.0){\shortstack{$\bullet$}}
\put(7.5,2.5){\shortstack{$\bullet$}}
\multiput(7.5,.5)(.0,3.0){2}{\shortstack{$\bullet$}}
\multiput(7.3,5.5)(0.0,3.0){2}{\shortstack{$\bullet$}}
\multiput(7.3,6.5)(.0,3.0){2}{\shortstack{$\bullet$}}
\put(7.05,.05){\line(1,0){1.0}}
\put(7.05,.05){\line(1,1){.5}}
\put(7.05,3.05){\line(1,1){.5}}
\put(7.55,2.5){\line(0,1){1.0}}
\put(7.35,5.5){\line(0,1){1.0}}
\put(7.35,8.5){\line(0,1){1.0}}
\put(4.0,-2.0){\shortstack{FIGURE 2}}
\multiput(9.0,0.0)(0.0,3.0){4}{\shortstack{$\bullet$}}
\put(-1.5,0.0){\shortstack{(d)}}
\put(-1.5,3.0){\shortstack{(c)}}
\put(-1.5,6.0){\shortstack{(b)}}
\put(-1.5,9.0){\shortstack{(a)}}
\end{picture}
\newpage
\begin{picture}(8,8)(-50,450)
\unitlength 1.5cm
\thicklines
\multiput(.0,.0)(0.0,3.0){3}{\shortstack{$\bullet$}}
\multiput(.5,.5)(.0,3.0){3}{\shortstack{$\bullet$}}
\multiput(.5,-.5)(.0,3.0){3}{\shortstack{$\bullet$}}
\multiput(1.0,.0)(.0,3.0){3}{\shortstack{$\bullet$}}
\multiput(2.5,.0)(.0,3.0){3}{\shortstack{$\bullet$}}
\multiput(3.0,.5)(.0,3.0){3}{\shortstack{$\bullet$}}
\multiput(3.0,-.5)(0.0,3.0){3}{\shortstack{$\bullet$}}
\multiput(3.5,.0)(0.0,3.0){3}{\shortstack{$\bullet$}}
\put(2.55,.05){\line(1,-1){.5}}
\put(3.05,3.55){\line(1,-1){.5}}
\put(2.55,6.05){\line(1,0){1.0}}
\put(3.05,5.55){\line(0,1){.4}}
\put(3.05,6.05){\line(0,1){.45}}
\put(-1.5,0.0){\shortstack{(c)}}
\put(-1.5,3.0){\shortstack{(b)} }
\put(-1.5,6.0){\shortstack{(a)}}
\put(3.0,-2.0){\shortstack{FIGURE 3(a)(b)(c)}}
\multiput(4.5,0.0)(.0,3.0){2}{\shortstack{$\bullet$}}
\multiput(5.0,.5)(.0,3.0){2}{\shortstack{$\bullet$}}
\put(5.5,.0){\shortstack{$\bullet$}}
\put(5.0,2.5){\shortstack{$\bullet$}}
\multiput(6.5,.0)(.0,3.0){2}{\shortstack{$\bullet$}}
\multiput(7.0,.5)(.0,3.0){2}{\shortstack{$\bullet$}}
\put(7.0,2.5){\shortstack{$\bullet$}}
\put(7.5,.0){\shortstack{$\bullet$}}
\put(4.8,6.0){\shortstack{$\bullet$}}
\multiput(8.5,0.0)(.0,3.0){2}{\shortstack{$\bullet$}}
\put(6.55,0.05){\line(1,0){1.0}}
\put(6.55,0.05){\line(1,1){.5}}
\put(6.55,3.05){\line(1,1){.5}}
\put(6.55,3.05){\line(1,-1){.5}}
\put(2.55,6.05){\line(1,1){.5}}
\put(3.05,5.55){\line(1,1){.5}}
\end{picture}

\newpage
\begin{picture}(8,8)(-60,450)
\unitlength 1.5cm
\thicklines
\put(0.0,0.0){\vector(0,1){6.0}}
\put(0.0,0.0){\vector(1,0){6.0}}
\put(2.0, 0.0){\line(0,1){1.0}}
\put(2.0,3.0){\line(0,1){2.0}}
\put(2.0,1.0){\line(1,0){3.0}}
\put(2.0,1.0){\line(0,1){2.0}}
\put(2.0,3.0){\line(1,2){1.0}}
\put(-0.5,2.0){\shortstack{2}}
\put(-0.5,4.0){\shortstack{4}}
\put(2.0,-.5){\shortstack{2}}
\put(4.0,-.5){\shortstack{4}}
\put(7.0,0.0){\shortstack{\Large$\gamma_0$}}
\put(0.0,7.0){\shortstack{\Large$\gamma_x$}}
\put(0.0,2.0){\shortstack{$-$}}
\put(0.0,4.0){\shortstack{$-$}}
\put(2.0,0.06){\shortstack{$|$}}
\put(4.0,.06){\shortstack{$|$}}
\put(0.6,2.5){\shortstack{ALL G}}
\put(3.0,.5){\shortstack{KONDO}}
\put(3.5,2.0){\shortstack{NFL}}
\put(2.3, 5.0){\shortstack{$G_2$}}
\put(2.5,-1.5){\shortstack{FIGURE 3(d)}}
\end{picture}

\newpage
\begin{picture}(8,8)(-60,450)
\unitlength 1.5cm
\put(0.0,0.0){\vector(0,1){6.0}}
\put(0.0,0.0){\vector(1,0){6.0}}
\put(2.0, 0.0){\line(0,1){1.0}}
\put(2.0,3.0){\line(0,1){2.0}}
\put(0.0,1.0){\line(1,0){5.0}}
\put(2.0,1.0){\line(0,1){2.0}}
\put(0.5,0.0){\line(1,2){2.2}}
\put(1.2,0.5){\shortstack{$I$}}
\put(1.2,3.0){\shortstack{$II$}}
\put(1.4,1.4){\shortstack{$III$}}
\put(0.2,.5){\shortstack{$III$}}
\put(-0.5,2.0){\shortstack{2}}
\put(-0.5,4.0){\shortstack{4}}
\put(2.0,-.5){\shortstack{2}}
\put(4.0,-.5){\shortstack{4}}
\put(3.0,-1.0){\shortstack{\Large$\gamma_0$}}
\put(-1.0,3.0){\shortstack{\Large$\gamma_x$}}
\put(0.0,2.0){\shortstack{$-$}}
\put(0.0,4.0){\shortstack{$-$}}
\put(4.0,.06){\shortstack{$|$}}
\put(3.0,.5){\shortstack{KONDO}}
\put(3.5,2.0){\shortstack{NFL}}
\put(2.3, 5.0){\shortstack{$G_2$}}
\put(2.5,-2.0){\shortstack{FIGURE 3(e)}}
\thicklines
\put(2.0, 1.0){\line(1,0){3.0}}
\put(2.0,1.0){\line(0,1){2.0}}
\put(2.0,3.0){\line(1,2){1.0}}
\end{picture}


\begin{references}

\bibitem{And1} P. W. Anderson, Science {\bf 235}, 1196(1987); Physica
{\bf C 185-189}, 11(1991); P. W. Anderson and R. Schrieffer,
Phys. Today, June 1991, p.54.
\bibitem{Var1} C. M. Varma, S. Schmitt-Rink and E. Abraham,
Solid Stat. Commun. {\bf 62}, 681(1987);
C. M. Varma, P. B. Littlewood, S. Schmitt-Rink,
E. Abraham and A. E. Ruckenstein, Phys. Rev. Lett. {\bf 63},
1996 (1989); A. E. Ruckenstein and C. M. Varma, Physica {\bf C 185-189},
134(1991).
\bibitem{Hald1} F. D. M. Haldane, J. Phys., {\bf C 14}, 2585 (1981).
\bibitem{Sh1} See,  e.g., R. Shankar, preprint cond-mat/9307009 and 
references therein.
\bibitem{Matt1} D. C. Mattis and E. Lieb, J. Math. Phys., {\bf 6},304(1965);
A. Luther and I. Peschel, Phys. Rev. {\bf B 12}, 3908 (1975);
S. Coleman, Phys. Rev. {\bf D 11}, 2088 (1975); S. Mandelstam,
Phys. Rev. {\bf D 11}, 3026 (1975).
\bibitem{Hald2} F. D. M. Haldane, Varenna Lecture, 1992 and Helv. Phys. Acta.
{\bf 65}, 152 (1992).
\bibitem{Hou1} A. Houghton and J. B. Marston, Phys. Rev. {\bf B 48},
7790 (1993).
\bibitem{Kh1} D. V. Khevshchenko, R. Hlubina and T. M. Rice, Phys. Rev.
{\bf B 48 }, 10766 (1993); A. H. Castro Neto and E. Fradkin,
Urbara preprint, 1993; R. Hlubina, ETH preprint,1993; D. V.
Khevshchenko, Princeton preprint, 1993.
\bibitem{And2} P. W. Anderson, Phys. Rev. Lett. {\bf 267}, 3844 (1991).
\bibitem{Wen1} X. G. Wen, Phys. Rev. {\bf B 42}, 6623 (1990);
H. J. Schulz, Int. J. Mod. Phys. {\bf B 5}, 57 (1991); M. Fabrizio,
A. Parola and E. Tosatti, Phys. Rev. {\bf B 46}, 3159 (1992);
G. M. Zhang, S. P. Feng and L. Yu, to be published;
A. Nersesyan, A. Luther and F. V. Kusmartev, Phys. Lett. {\bf A
176}, 363 (1993); P. W. Anderson, JETP Lett. {\bf 58}, 59 (1993);
V. M. Yakovenko, {\it ibid} {\bf 56}, 5101 (1992); D. V. Khevshchenko,
preprint; Y. M. Li, Y. Yu, N. d'Amberumenil, L. Yu and Z. B. Su, 
Mod. Phys. Lett. {\bf B 8},749 (1994); and
references therein.
\bibitem{Fri1} J. Friedel, Can. J. Phys. {\bf 34}, 1190 (1956);
A. Blandin and J. Friedel, J. Phys. Radium {\bf 20}, 160 (1959).
\bibitem{And3} P. W. Anderson, Phys. Rev. {\bf 124}, 41(1961).
\bibitem{Wil1} K. G. Wilson, Rev. Mod. Phys. {\bf 47}, 773(1975).
\bibitem{Noz1} P. Nozieres, J. Low Temp. Phys. {\bf 17}, 31(1974).
\bibitem{Per1} I. E. Perakis, C. M. Varma and A. E. Ruckenstein,
Phys. Rev. Lett. {\bf 70}, 3467 (1993).
\bibitem{Gia1} T. Giamarchi, C. M. Varma,
A. E. Ruckenstein and P. Nozieres, Phys. Rev. Lett. {\bf 70},
3967 (1993).
\bibitem{Zha1} G. M. Zhang, L. Yu and Z. B. Su, Phys. Rev. {\bf B 49}, 
7759(1994).
\bibitem{Si1} Q. M. Si and G. Kotliar, Phys. Rev. Lett. {\bf 70},
3143 (1993).
\bibitem{Wie1} P. B. Wiegmann and A. M. Finkel'stein, Sov. Phys. JETP
{\bf 48}, 102 (1978); P. Schlottmann, J. Phys. (Paris) {\bf 6}, 1486(1978);
Phys. Rev. {\bf  B 45}, 4815 (1982).
\bibitem{Sol1} J. Solyom, Adv. Phys. {\bf 28}, 209 (1979); J. Phys.
{\bf F 4}, 2269 (1975); P. Nozieres,and A. Blandin, J. Phys.
(Paris) {\bf 41}, 193 (1980).
\bibitem{And4} P. W. Anderson, G. Yuval and D. R. Hamann, Phys. Rev.
{\bf B1}, 4464 (1970); P. W. Anderson, J. Phys. {\bf C3}, 2436 (1970).
\bibitem{Mah1} G. Mahan, Phys. Rev. {\bf 153}, 882 (1967);
P. W. Anderson, Phys. Rev. Lett. {\bf 18}, 1049 (1967);
P. Nozieres and C. T. de Dominicis, Phys. Rev. {\bf 173}, 1097 (1969).
\bibitem{Yu1} Yue Yu, Y.M. Li and N. d'Amberumenil,
Phys. Rev. {\bf B 51}, 16417 (1995).
\bibitem{Car1} J. L. Cardy, J. Phys. {\bf A14}, 1407 (1981); See also
S. Chakravarty and J. Hirsch, Phys. Rev. {\bf B25}, 3273 (1982).
\bibitem{SK} Q. Si and G. Kotliar, Phys. Rev. {\bf B48}, 13881 (1993).
\bibitem{Eme1} V. J. Emery, in {\it Highly Conducting One-dimensional
Solids}, eds. J. T. Devreese  {\it et al}. (plenum, New York,1979);
J. Solyom in \cite{And4}.
K. D. Schotte and U. Schotte, Phys. Rev. {\bf 182},479 (1969); K. D.
Schotte, Z. Phys. {\bf 230}, 99 (1970).
\bibitem{Sir1} C. Sire, C. M. Varma, A. E. Ruckenstein and T. Giamarchi,
Phys. Rev. Lett.{\bf 72},2478 (1994).
\bibitem{Zha2} G. M. Zhang and L. Yu, Phys. Rev. Lett. {\bf 72}, 2474 (1994).
\bibitem{27} In \cite{Per1}, this term was discussed as the leading term of the
fixed point Hamiltonian in the second order.
\bibitem{Hald3} F. M. D. Haldane, Phys. Rev. Lett. {\bf 40}, 416 (1978);
J. Phys. {\bf C 11}, 5015 (1978).
\bibitem{29} In the language of conformal field theory, $O(\alpha,\beta)$
is called vertex operator; $q^a_{\alpha\beta}$ is still called  charge; and
$K(\alpha,\beta)$ is the conformal weight of the vertex operator
$O(\alpha,\beta)$.
\bibitem{Wie2} See P. B. Wiegmann and A. M. Finkel'stein in \cite{Wie1}.
\bibitem{32} See, for example, P. Nozieres and C. T. de Dominicis in 
\cite{Mah1};
Also see K. D. Schotte and  U. Schotte in \cite{Eme1}.
Recent application, see \cite{Per1}.

\end{references}
\end{document}